\documentclass[11pt,reqno]{article}
\usepackage{graphicx,amsmath,amssymb,setspace}
\usepackage{natbib}

\newtheorem{remark}{Remark}[section]

\numberwithin{equation}{section}

\newcommand{\eps}{\epsilon}
\renewcommand{\l}{\mathcal{L}}
\newcommand{\tl}{\widetilde{\mathcal{L}}}
\newcommand{\m}{\mathcal{M}}
\newcommand{\tm}{\widetilde{\mathcal{M}}}
\newcommand{\tc}{\tilde{c}}
\newcommand{\tg}{\tilde{g}}
\newcommand{\tv}{\tilde{v}}

\newcommand{\tphi}{\tilde{\phi}}
\newcommand{\tL}{\widetilde{\Lambda}}
\newcommand{\La}{\Lambda}
\newcommand{\Ga}{\Gamma}
\newcommand{\tG}{\widetilde{\Gamma}}
\newcommand{\tV}{\widetilde{V}}

\newcommand{\tB}{\widetilde{B}}

\title{\LARGE \bf
\sc Pricing Options on Defaultable Stocks \thanks{This
work is supported in part by the National Science Foundation,
under grant DMS-0604491.} \thanks{This work was inspired by the SAMSI
workshops on Financial Mathematics, Statistics and Econometrics
(Fall 2005, Spring 2006 North Carolina). The author wishes to thank the organizers
for the travel grant to participate in this stimulating event. I also would like to thank Bo Yang for his research assistance and the two anonymous referees and an anonymous associate editor for their valuable suggestions.}}

\author{Erhan Bayraktar 
\thanks{Department of Mathematics, University
of Michigan, Ann Arbor, MI 48109, USA, email: erhan@umich.edu} }
\date{}

\begin{document}
\maketitle

\begin{abstract}\noindent
We develop stock option price approximations for a model which takes both the risk of default and the stochastic volatility into account. We also let the intensity of defaults be influenced by the volatility. We show that it might be possible to infer the risk neutral default intensity from the stock option prices.
Our option price approximation has a rich implied volatility surface structure and fits the data implied volatility well. Our calibration exercise shows that an effective hazard rate from bonds issued by a company can be used to explain the implied volatility skew of the option prices issued by the same company. We also observe that the implied yield spread that is obtained from calibrating all the model parameters to the option prices matches the observed yield spread.

\end{abstract}

\section{Introduction}

Bonds and stocks issued by the same company carry the risk of the
company going bankrupt. Although this risk has been accounted for
in pricing the bonds, see e.g. \cite{schoenbucher} for a review of this literature, there is only
a limited number of models in equity option pricing, \cite{carr-linetsky}, \cite{carr-wu}, \cite{hull-nelken} and \cite{Linetsky}, \cite{merton-jump}.
Instead, models are developed to match the implied volatility
curve, which is observed daily: stochastic volatility models, see
e.g. \cite{sircar}, and jump diffusion models, see e.g.
\cite{cont}. These models account for the default risk implicitly.
These approaches could be improved accounting for the risk of default.
When a company goes
bankrupt and the stock of the company plunges, the out-of-the-money put options are suddenly very valuable. So, when these contracts
are priced, the market must also be accounting for these crash
scenarios, besides the stochastic volatility and small jump effects, and this is partly why the out of the money put options are worth
much more than what the Black and Scholes model, which does not account
for bankruptcy, would predict.

In addition, one might argue that the default risk should be
priced consistently across the markets. When the likelihood of the
default of a company increases (for example when a credit
downgrade occurs), it is not unreasonable to expect that besides
the increase in the yield spread, the trading volume in the out of
the money put options increases and their prices go up. In fact,
the more the put option is out of the money, the higher the return it
will bring when there is a default, so the higher the prices
should be. This is the understanding of \cite{Linetsky} and the
convertible bond pricing literature(see e.g. \cite{vetzal}).

In this note, we will use an intensity based approach to model the
default risk of the company that issues the stock. We will also allow the volatility of the stock price to be stochastic. We observe that the implied yield spread that is obtained from calibrating all the model parameters to the option prices matches the observed yield spread (Figure 15). On the other hand, our calibration exercise shows that an effective hazard rate from bonds issued by a company can be used to explain the implied volatility skew of the option prices issued by the same company. 

In our model, the volatility is driven by two processes that evolve on two different time scales, fast and slow. The intensity of default is a function of volatility and an idiosyncratic stochastic component (since the credit market and stock options market can have independent movements), which is driven by two other  processes that evolve on fast and slow time scales. Our model is similar to that of \cite{carr-wu} and the motivation for our model specification can be found on page 4 of that paper. The difference of our model from the Carr-Wu specification is that we also incorporate the market prices of volatility and intensity risks into our model, and we consider the intensity and volatility to be evolving on two different time scales. The purpose of our specification is to be able to obtain asymptotic option price approximations using multiscale perturbation methods developed by \cite{ronnie-timescale}, who used this
technique in the context of stochastic volatility models. Our model can be thought of as a hybrid of the stochastic volatility model of \cite{ronnie-timescale} (whose purpose is to obtain approximations for stock option prices) and the multi-scale intensity model of \cite{papa} (whose purpose is to price credit derivatives). But one should note that  following the Carr-Wu specification we allow the intensity to depend on the volatility, and we also incorporate the market price of default risk into our model.
Let us also point out that our default model differs from \cite{carr-linetsky} and \cite{Linetsky} since they
consider a model in which the intensity of the defaults depends solely on the
stock price.

Our price approximation for the option prices has seven parameters (we will refer to it as the 7 parameter model), one of which is the average intensity, that can be calibrated to the observed implied volatility. We compare the performance of this model against 3 simpler models: [5 parameter model] a model that does not account for the market price of intensity risk; [the model of \cite{ronnie-timescale}] a model that does not account for the default risk; [3 parameter model] a model that does not account for the stochastic volatility effects.
When we take the average intensity to be equal to the yield spread of the corporate bond with smallest maturity and calibrate the rest of the parameters to the observed option prices we observe that the seven parameter model not only outperforms the others (which is to be expected) but also that it almost fits the data implied volatility perfectly for longer term options (9 months and over). (In our calibration we pool the options with different strikes and different maturities in the same pool and we have 104 data points, we consider maturities from 3 months to 2 years). The 5 parameter model always outperforms the model of \cite{ronnie-timescale}, which points to the significance of accounting for the default risk.
Although, the three parameter model has only two free parameters (the other free parameter is the average intensity of default and is already fixed to be the yield spread of the shortest maturity bond) it fits the implied volatility surface fairly well, considering that we have 104 data points. See Section 3.2.

\begin{figure}[h]
\begin{center}
\includegraphics[width = 0.85\textwidth,height=6cm]{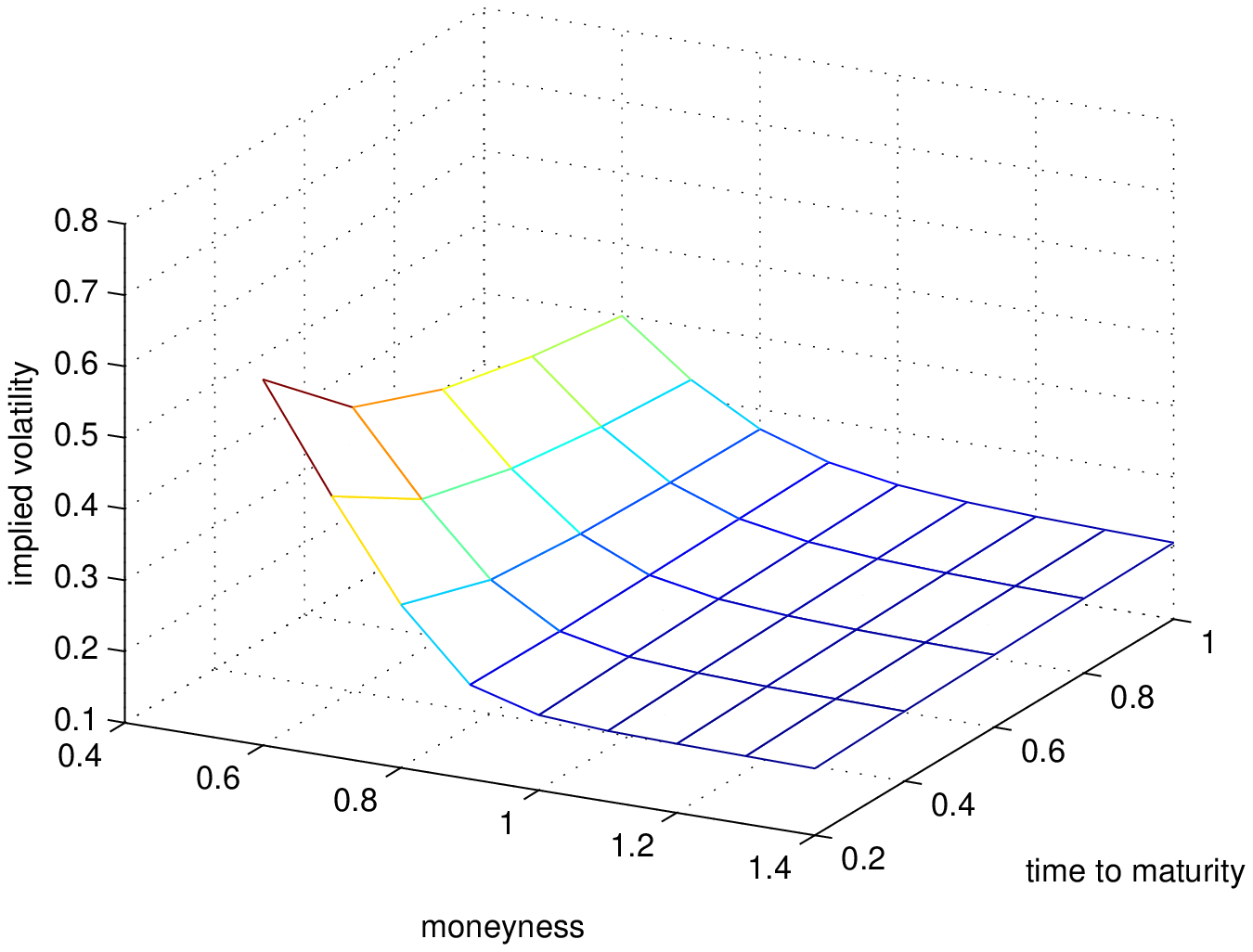}
\caption{The Implied Volatility Surface of the 3 parameter model in (\ref{eq:comp-mod-2}). Average volatility=0.2, average intensity=$\bar{\lambda}=0.02$, interest rate(r)=0.04 (assuming $B(t,T)=e^{r(T-s)}$), $V^{\eps}=0.0015$, and $V^{\delta}=0.001$}
\label{fig:s3p}
\end{center}
\end{figure}

\begin{figure}[h]
\begin{center}
\includegraphics[width = 0.85\textwidth,height=6cm]{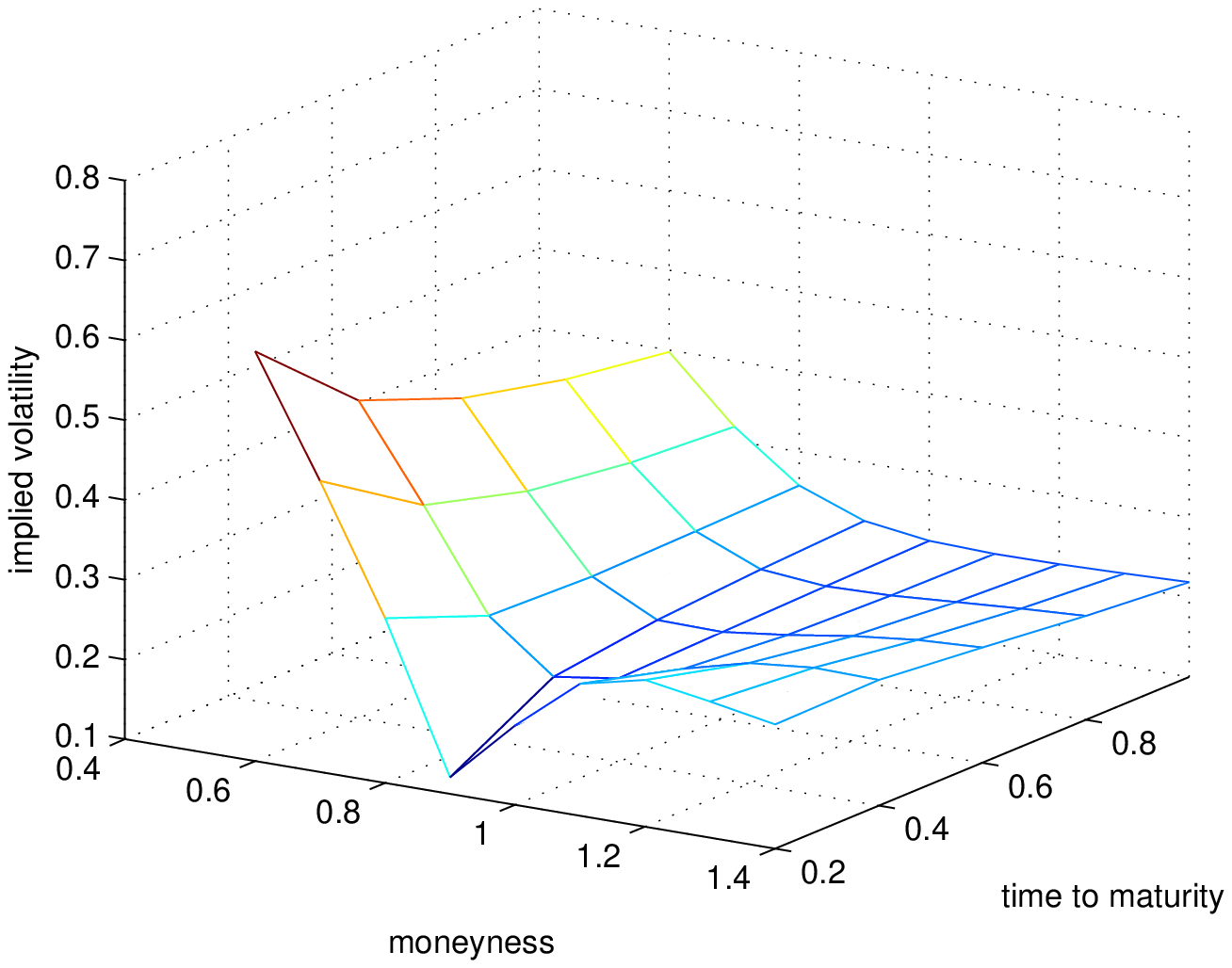}
\caption{The Implied Volatility Surface corresponding to the 5 parameter model in (\ref{eq:comp-mod-1}). Average volatility=0.2, $\bar{\lambda}=0.02$, r=0.04,  $\tV_1^{\eps}=-0.0015$ $\tV_2^{\eps}=0.001$, $\tV_1^{\delta}=-0.001$ $\tV^{\delta}=-0.001$.}
\label{fig:s5p}
\end{center}
\end{figure}

When we calibrate all parameters of each model, including the average intensity, to the option prices then all the models we consider (except the model of \cite{ronnie-timescale} since this model does not account for the default risk) fit the the data implied volatility curve quite well (even the 3 parameter model). But when option implied average intensity is plotted against the yield curve of the shortest maturity bond over the entire period of the data (about a year), one observes that all models except the 7 parameter model have unrealistically severe option implied intensities. Only the 7 parameter model almost matches the yield spread of the shortest maturity bond (the maturity is about two years on the last day of our data). Therefore, it might be possible to derive the risk neutral probability of default from the option prices using the 7 parameter model. See Section 3.3. Also we should note that the 7 and 5 parameter models have much richer implied volatility surface structure (see Figures~\ref{fig:s3p}, \ref{fig:s5p}, \ref{fig:sfp}, and \ref{fig:s7p}) and Section 2.3, which explains why they fit to the option prices better.

We should note that we calibrate the parameters of our model to the daily observed option prices as in \cite{ronnie-timescale} and \cite{papa}.  \cite{carr-wu} on the other hand perform a time series analysis to obtain the parameters of their model.

The rest of the paper is organized as follows: In Section 2, we
introduce our model specification. In Section~\ref{sec:eop}, we derive an
approximation of the prices of the equity options, and in Section 2.2 we derive an approximation to the corporate bond prices, using multi-scale perturbation techniques. In Section 2.3 we analyze the model implied volatility
surface structure of the models we consider. In Section 3, we calibrate the 7 parameter model and the other reference models to the data.

\section{A Multiscale Intensity Model}

We will use the doubly-stochastic Poisson process framework of
\cite{bremaud}, see also \cite{schoenbucher}. Let $(\Omega,
\mathcal{H},\mathbb{P})$ be a filtered probability space hosting standard
Brownian motions $\vec{W}=(W_t^{(0)}, W_t^{(1)}, W_t^{(2)}, W_t^{(3)}, W_t^{(4)})$, whose correlation structure is given by (2.8)
and standard Poisson process $N=(N(t))_{t \geq 0}$ independent of
$\vec{W}$. These Brownian motion will be used in modeling the the volatility and the intensity (see (2.6)).We model the time of default, which we
will denote by $\tau$, as the first time the process the time changed Poisson process $\tilde{N}_t
\triangleq N\left(\int_0^{t}\lambda_s ds\right)$, $t \geq 0$,
jumps.

The stock price dynamics in our framework is the solution of the
stochastic differential equation
\begin{equation}
d\tilde{S}_t=\tilde{S}_t\left(r_t dt+\sigma_t
dW^{(0)}_t-d\left(\tilde{N}_t-\int_0^{t \wedge \tau}\lambda_u
du\right)\right), \quad \tilde{S}_0=x,
\end{equation}
in which we take the interest rate $(r_t)_{t \geq 0}$ to be deterministic as Carr and Wu (2006) and Linetsky (2006) do.
At the time of default the stock price jumps down to zero. Note
that the discounted stock price is a martingale under the measure
$\mathbb{P}$. Let $(\mathcal{F}_t)_{t \geq 0}$ be the natural filtration of $\vec{W}$.
The price of a European call option with maturity $T>0$ is equal
to
\begin{equation}\label{eq:price-option}
\begin{split}
C(t)=\mathbb{E}\left[B(t,T)(\tilde{S}_T-K)^+1_{\{\tau>T\}}
\bigg|\mathcal{F}_t \vee \sigma\{\tilde{N}_s; 0 \leq s \leq t\}\right]=\mathbb{E}\left[B(t,T) e^{-\int_t^{T}\lambda_s
ds}(S_T-K)^+\bigg|\mathcal{F}_t\right],
\end{split}
\end{equation}
in which $B(t,T)$ is the price of the bond that matures at time $T$, whereas the price of a European put option is given by
\begin{equation}\label{eq:price-option-put}
\begin{split}
P(t)&=\mathbb{E}\left[B(t,T)((K-\tilde{S}_T)^+1_{\{\tau>T\}}+K1_{\{\tau
\leq T\}}) \bigg|\mathcal{F}_t \vee \sigma\{\tilde{N}_s; 0 \leq s \leq t\}\right]
\\&=KB(t,T)-\mathbb{E}\left[B(t,T)e^{-\int_t^{T}\lambda_s
ds}\min(K,S_T)\bigg|\mathcal{F}_t\right],
\end{split}
\end{equation}
in which $S=(S_t)_{t \geq 0}$ is the solution of
\begin{equation}\label{eq:dyn-stock}
dS_t=S_t(r+\lambda_t)+S_t \sigma_t dW^{(0)}_t, \quad S_0=x.
\end{equation}
From (\ref{eq:price-option}) and (\ref{eq:dyn-stock}) we see that
pricing a call option on a defaultable stock is equivalent to
pricing a call option in a fictitious market that does not default
but whose interest rate is stochastic (also see \cite{Linetsky}).
Using (\ref{eq:price-option}) and (\ref{eq:price-option-put}), the
put-call parity is satisfied,
\begin{equation}
C(t)-P(t)=S_t-B(t,T)K,
\end{equation}
as expected.

We will use a model specification similar to that of \cite{carr-wu} and assume the following dynamics for the volatility $\sigma_t$ and the intensity $\lambda_t$:
\begin{equation}\label{eq:model}
\begin{split}
\sigma_t&=\sigma(Y_t,Z_t), \quad \lambda_t=\beta \sigma_t^2+f(Q_t, U_t),
\end{split}
\end{equation}
in which
\begin{equation}
\begin{split}
dY_t&=\left(\frac{1}{\epsilon}(m-Y_t)-\frac{v \sqrt{2}}{\sqrt{\eps}}\Lambda(Y_t,Z_t)\right)dt+\frac{v \sqrt{2}}{\sqrt{\epsilon}}dW_t^{(1)}, \quad Y_0=y,
\\ dZ_t&=\left(\delta c(Z_t)-\sqrt{\delta}g(Z_t)\Gamma(Y_t,Z_t)\right)dt+\sqrt{\delta}g(Z_t)dW_t^{(2)}, \quad Z_0=z,
\\dQ_t&=\left(\frac{1}{\epsilon}(\tilde{m}-Q_t)-\frac{\tv \sqrt{2}}{\sqrt{\eps}}\tL(Q_t,U_t)\right)dt+\frac{\tilde{v}
\sqrt{2}}{\sqrt{\epsilon}}dW_t^{(3)}, \quad Q_0=q,
\\  dU_t&=\left(\delta \tc(U_t)-\sqrt{\delta}\tg(Z_t)\tG(Q_t,U_t)\right)dt+\sqrt{\delta}\tg(U_t)dW_t^{(4)}, \quad U_0=u.
\end{split}
\end{equation}
 Here, $\La$ and $\Ga$ are the market prices of volatility risk,
$\tL$ and $\tG$ are the market prices of intensity risk.
We will assume the following correlation structure for the Brownian motions in our model:
\begin{equation}
\begin{split}
E[W^{(0)}_t,W^{(i)}_t]&=\rho_i t ,\quad i=1,  2,\\
E[W^{(1)}_t,W^{(2)}_t]&=\rho_{12} t, \quad E[W^{(3)}_t,W^{(4)}_t]=\rho_{34} t, \\
E[W^{(i)}_t,W^{(3)}_t]&=0, \quad i=0, 1, 2, \quad  E[W^{(i)}_t,W^{(4)}_t]=0, \quad i=0, 1, 2,
\end{split}
\end{equation}
for all $t \geq 0$. This model captures the covariation of the implied volatility and the spread of risky bonds, and it also accommodates the fact that the credit market and the stock market can show independent movements (see Section 1.1 of \cite{carr-wu} for further motivation). One should note that when $\beta=0$, our model specification for the intensity coincides with that of \cite{papa}. On the other hand, when $\lambda=0$ our model specification becomes that of \cite{ronnie-timescale}. We assume that the market price of risks are bounded functions as in  \cite{ronnie-timescale}.

As in \cite{ronnie-timescale} and \cite{papa} we assume that both $\sigma$ and $f$ are  bounded, smooth, and strictly positive
functions. We also assume that the
differential equation defining $Z$ and $U$ have unique strong solutions.
We should also note that we think of $\varepsilon>0$, $\delta>0$
as small constants. Hence, the process $Y$ and $Q$ are fast mean reverting
and $Z$ and $U$ evolve on a slower time scale. See
\cite{ronnie-timescale} for an exposition of multi-scale modeling
in the context of stochastic volatility models.

\subsection{European Option Price} \label{sec:eop}

From an application of the Feynman-Kac Theorem it follows that the
price of a call option satisfies the following partial
differential equation
\begin{equation}\label{eq:option-pricing-pde}
\begin{split}
\l^{\eps,\delta}C^{\eps,\delta}(t,x,y,z,q,u)&=0, \quad t <T,
\\ C^{\eps,\delta}(T,x,y,z,q,u)&=(x-KB(t,T))^+,
\end{split}
\end{equation}
where the operator $\l^{\eps,\delta}$ is given by
\begin{equation}
\l^{\eps,\delta}\triangleq \frac{1}{\eps}
\l_0+\frac{1}{\sqrt{\eps}}\l_1+\l_2+\sqrt{\delta}\m_1+\delta \m_2+
\sqrt{\frac{\delta}{\eps}}\m_3,
\end{equation}
in which
\begin{equation}\label{eq:operators}
\begin{split}
\l_0 & \triangleq(m-y)\frac{\partial}{\partial y}+v^2\frac{\partial^2}{\partial y^2} +(\tilde{m}-q)\frac{\partial}{\partial q}+\tilde{v}^2\frac{\partial^2}{\partial q^2},\\
\l_1 &\triangleq  (\rho_1\sigma(y,z)v\sqrt{2})x\frac{\partial^2}{\partial x \partial y}-\La(y,z) v \sqrt{2} \frac{\partial}{\partial y}-\tL(q,u)\tv \sqrt{2} \frac{\partial}{\partial q},\\
\l_2 &\triangleq \frac{\partial}{\partial t}+\frac{1}{2}\sigma^2(y,z)x^2\frac{\partial}{\partial x^2}+(\beta \sigma^2(y,z)+f(q,u))x\frac{\partial}{\partial x}-(\beta \sigma^2(y,z)+f(q,u))\cdot,\\
\m_1 &\triangleq  \rho_2\sigma(y,z)g(z)x\frac{\partial^2}{\partial x \partial z}-g(z)\Ga(y,z)\frac{\partial}{\partial z}-\tg(z)\tG(q,u)\frac{\partial}{\partial u},\\
\m_2 &\triangleq  c(z)\frac{\partial}{\partial z}+\frac{1}{2}(g(z))^2\frac{\partial^2}{\partial z^2}+\tilde{c}(z)\frac{\partial}{\partial z}+\frac{1}{2}(\tg(z))^2\frac{\partial^2}{\partial z^2},\\
\m_3 &\triangleq \rho_{12}v\sqrt{2}g(z)\frac{\partial^2}{\partial y
\partial z}+\rho_{34}\tilde{v}\sqrt{2}\tg(z)\frac{\partial^2}{\partial q
\partial z}.
\end{split}
\end{equation}
We have denoted the price of a call option by $C^{\eps,\delta}$ to
emphasize the dependence on the parameters $\eps$ and $\delta$.

Here, the operator $\l_2$ is the Black-Scholes operator
corresponding to interest rate level $\beta \sigma^2(y,z)+f(q,u)$ and volatility $\sigma(y,z)$.
$(1/\eps) \l_0$ is the infinitesimal generator of the
Ornstein-Uhlenbeck process $(Y,Q)$. $\l_1$ contains the mixed partial
derivative due to the correlation between the Brownian motions
driving the stock price and the fast volatility factor. $\m_1$ contains the
mixed partial derivative due to the correlation between the
Brownian motions driving the stock price and the slow volatility factor.
$\delta \m_2$ is the infinitesimal generator of the two-dimensional process $(Z,U)$. $\m_3$
contains the mixed partial derivatives due to the correlation
between the Brownian motions driving the fast and slow factors of
the volatility and the intensity.

\noindent \textbf{Approximating the Price Using the Multiscale Perturbation Method}

We will first consider an expansion of the price $C^{\eps,\delta}$
in powers of $\sqrt{\delta}$,
\begin{equation}\label{eq:delta-expansion}
C^{\eps,\delta}=C_{0}^{\eps}+\sqrt{\delta}C_1^{\eps}+\delta
C_2^{\eps}+\cdots,
\end{equation}
and then consider the expansion of each term in
(\ref{eq:delta-expansion}) in powers of $\sqrt{\eps}$, i.e.,
\begin{equation}\label{eq:expansion-eps}
C_k^{\eps}=C_{0,k}+\sqrt{\eps}C_{1,k}+\eps C_{2,k}+ \cdots, \quad
\text{for all $k \in \mathbb{N}$}.
\end{equation}
It can be seen from (\ref{eq:option-pricing-pde}) and
(\ref{eq:delta-expansion}) that $C^{\eps}_0$ and $C^{\eps}_1$
satisfy
\begin{equation}\label{eq:C-0-eps}
\begin{split}
\left(\frac{1}{\eps}\l_0+\frac{1}{\sqrt{\eps}}\l_1+\l_2\right)C^{\eps}_0&=0,
\\ C^{\eps}_0(T,x,y,z)&=(x-K)^+, \quad \text{and}
\end{split}
\end{equation}
\begin{equation}\label{eq:C-1-eps}
\begin{split}
\left(\frac{1}{\eps}\l_0+\frac{1}{\sqrt{\eps}}\l_1+\l_2\right)C^{\eps}_1&=-\left(\m_1+\frac{1}{\sqrt{\eps}}\m_3\right)C_{0}^{\eps},
\\ C_1^{\eps}(T,x,y,z)&=0.
\end{split}
\end{equation}
Equating the $1/\eps$ and $1/\sqrt{\eps}$ terms in
(\ref{eq:C-0-eps}), we obtain
\begin{equation}\label{eq:eps-sqrt-eps}
\l_0 C_{0,0} =0, \quad \l_0 C_{1,0}+ \l_1 C_{0,0}=0.
\end{equation}
Since the first equation in (\ref{eq:eps-sqrt-eps}) is a
homogeneous equation in $y$ and $q$, we take $C_{0,0}=C_{0,0}(t,x,z,u)$ to be
independent of $y$ and $q$. (The $y$ and $q$ dependent solutions have exponential growth at infinity (\cite{sircar-proof}).) Similarly, from the second equation, it can be
seen that $C_{1,0}$ can be taken to be independent of $y$ and $q$, since $\l_1 C_{0,0}=0$.
When we equate the order one terms and use the fact that $\l_1
C_{1,0}=0$, we get
\begin{equation}\label{eq:Poisson-eq-1}
\l_0C_{2,0}+\l_2 C_{0,0}=0,
\end{equation}
which is known as a Poisson equation (for $C_{2,0}$). (See
\cite{sircar}.) The solvability condition of this equation
requires that
\begin{equation}\label{eq:Bs-eqn-average-int-rate}
\left<\l_2\right>C_{0,0}=0, \quad \text{in which}
\end{equation}
\begin{equation}
\left<\l_2\right> \triangleq \frac{\partial}{\partial t}+\frac{1}{2}\left<\sigma^2(y,z)\right>x^2\frac{\partial}{\partial x^2}+(\beta \left<\sigma^2(y,z)\right>+\left<f(q,u)\right>)x\frac{\partial}{\partial x}-(\beta \left<\sigma^2(y,z)\right>+\left<f(q,u)\right>)\cdot,
\end{equation}
where $\left<\sigma^2(y,z)\right>$, which is only a function of $z$,
is the average (expectation) with respect to
the invariant distribution of $Y$, which is Gaussian with mean $m$
and standard deviation $\nu$, and $\left<f(q,u)\right>$, which is only a function of $u$, is the average with respect to the invariant distribution of $Q$,
which is Gaussian with mean $\tilde{m}$ and standard deviation $\tilde{\nu}$. All of this is equivalent to saying that the average $\left<\l_2\right>$ of the Black-Scholes operator $\l_2$ is with respect to the invariant distribution of $(Y,Q)$, whose density is given by
\begin{gather}
\Psi(y,q) \triangleq \frac{1}{2\pi v\tilde
v}\exp\left\{-\frac{1}{2}\left[\left(\frac{y-m}{v}\right)^2+\left(\frac{q-\tilde{m}}{\tilde{v}}\right)^2\right]\right\}.
\end{gather}

The solution to (\ref{eq:Bs-eqn-average-int-rate}) is
the Black-Scholes call option price when the volatility and the interest rate are
taken to be $\left<\sigma^2(y,z)\right>$ and $\beta \left<\sigma^2(y,z)\right>+\left<f(q,u)\right>$,
\begin{equation}\label{BS-formula}
\begin{split}
C_{0,0}(t,x,z,u)&= C_{\text{BS}}\left(x; \sqrt{\left<\sigma^2(y,z)\right>},
\beta \left<\sigma^2(y,z)\right>+\left<f(q,u)\right>;K B(t,T),T-t\right) \\ & \triangleq x
N(\tilde{d}_1)-KB(t,T)\exp\left(-(\beta \left<\sigma^2(y,z)\right>+\left<f(q,u)\right>)(T-t)\right)N(\tilde{d}_2),
\end{split}
\end{equation}
where
\begin{equation}\label{eq:d-1-tilde}
\tilde{d}_{1,2}=\frac{\log(x/[K B(t,T)])+(\beta \left<\sigma^2(y,z)\right>+\left<f(q,u)\right>\pm\frac{1}{2}
\left<\sigma^2(y,z)\right>)(T-t)}{\sqrt{\left<\sigma^2(y,z)\right>
(T-t)}}.
\end{equation}

From (\ref{eq:Poisson-eq-1}) and
(\ref{eq:Bs-eqn-average-int-rate}), we obtain that
\begin{equation}
C_{2,0}=-\l^{-1}_0 \left(\l_2-\left<\l_2\right>\right)C_{0,0}.
\end{equation}
 Equating
the order $\sqrt{\eps}$ terms in (\ref{eq:C-0-eps}) gives
\begin{equation}
\l_0 C_{3,0}+ \l_1 C_{2,0}+ \l_2 C_{1,0}=0,
\end{equation}
which is again a Poisson equation for $C_{3,0}$ and its
solvability condition requires that
\begin{equation}\label{eq:av-l2c-10}
\left<\l_2\right>C_{1,0}=-\left<\l_1 C_{2,0}\right>=\left<\l_1 \l_0^{-1}(\l_2-\left<\l_2\right>)\right>C_{0,0}.
\end{equation}
Observe that we have $C_{1,0}(T,x,z,u)=0$. We can compute $C_{1,0}$ explicitly. To facilitate this calculation we will need to introduce two new functions. Let $\phi(y,z)$ and $\tphi(q,u)$ be the solutions of
\begin{equation}
\l_0\phi(y,z,q,u)=\sigma^2(y,z)-\left<\sigma^2(y,z)\right> \quad \text{and} \quad \l_0 \tphi(y,z,q,u)=f(q,u)-\left<f(q,u)\right>,
\end{equation}
respectively. Now, one can easily deduce that
\begin{equation}
\l_0^{-1}(\l_2-\left<\l_2\right>)C_{0,0}=\phi\cdot\left(\frac{1}{2}x^2 \frac{\partial C_{0,0}}{\partial x^2}+\beta x \frac{\partial C_{0,0}}{\partial x}-\beta C_{0,0}\right)+\tphi\cdot\left(x \frac{\partial C_{0,0}}{\partial x}-C_{0,0}\right).
\end{equation}
On the other hand, using (\ref{eq:av-l2c-10}) it can be seen that
\begin{equation}\label{eq:l-2-c-1-0}
\begin{split}
&\left<\l_2\right>C_{1,0}=
\left(\frac{1}{\sqrt{2}}v\rho_1\langle \sigma \phi_y\rangle\right)x\frac{\partial}{\partial x}(x^2\frac{\partial^2C_{0,0}}{\partial x^2})
+\beta(\sqrt{2}v\rho_1\langle \sigma \phi_y\rangle)x^2\frac{\partial^2 C_{0,0}}{\partial x^2}+(\sqrt{2}v\rho_1\langle \sigma \tilde{\phi}_y\rangle)x^2\frac{\partial^2 C_{0,0}}{\partial x^2}
\\&- v \sqrt{2}\left<\phi_y \La\right>\left(\frac{1}{2}x^2 \frac{\partial C_{0,0}}{\partial x^2}+\beta x \frac{\partial C_{0,0}}{\partial x}-\beta C_{0,0}\right)-v \sqrt{2}\left<\tphi_y \La\right>\left(x \frac{\partial C_{0,0}}{\partial x}-C_{0,0}\right)
\\&-\tv \sqrt{2}\left<\tL \phi_q\right>\left(\frac{1}{2}x^2 \frac{\partial C_{0,0}}{\partial x^2}+\beta x \frac{\partial C_{0,0}}{\partial x}-\beta C_{0,0}\right)-\tv \sqrt{2}\left<\tL \tphi_q\right>\left(x \frac{\partial C_{0,0}}{\partial x}-C_{0,0}\right)
\end{split}
\end{equation}
in which $\phi_y$ and $\tphi_y$ are the derivatives of $\phi$ and $\tphi$, respectively, with respect to the $y$ variable.
The solution of (\ref{eq:l-2-c-1-0}) can be explicitly computed as
\begin{equation}
C_{1,0}=-(T-t)\text{RHS},
\end{equation}
in which $\text{RHS}$ is the right-hand-side of (\ref{eq:l-2-c-1-0}),
using the fact that $\left<\l_2\right>$ and
$x^n\partial^n/\partial x^n$ commute for $n\in \mathbb{N}$.

Next, we consider (\ref{eq:C-1-eps}) together with
(\ref{eq:expansion-eps}). Equating $1/\eps$ and $1/\sqrt{\eps}$
order terms gives
\begin{equation}\label{eq:1-over-epsilon}
\l_0 C_{0,1}=0, \quad \l_0 C_{1,1}=0.
\end{equation}
The first equation in (\ref{eq:1-over-epsilon}) implies that
$C_{0,1}$ is independent of $y$ and $q$, and the second equation can be
obtained by observing that $\m_3 C_{0,0}=\l_1 C_{0,1}=0$, which
implies that $C_{1,1}$ is independent of $y$ and $q$ as well.

Matching the first order terms gives
\begin{equation}\label{eq:poisson-eqn-2}
\l_0 C_{2,1}+ \l_2 C_{0,1}=-\m_1 C_{0,0},
\end{equation}
where we have used the fact that $\m_3 C_{1,0}=\l_1 C_{1,1}=0$.
Equation (\ref{eq:poisson-eqn-2}) is a  Poisson equation and its
solvability condition gives
\begin{equation}\label{eq:c10}
\left<\l_2\right>C_{0,1}=-\left<\m_1\right> C_{0,0}, \quad C_{0,1}(T,x,y,z)=0.
\end{equation}
Using the facts that
\begin{equation}
\begin{split}
\frac{\partial}{\partial z}C_{0,0}&=(T-t)\left<\sigma^2(y,z)\right>_{z}\left( \frac{1}{2}x^2 \frac{\partial^2 C_{0,0}}{\partial x^2}+\beta x \frac{\partial}{\partial x}C_{0,0}-\beta C_{0,0}\right),
\\ \frac{\partial}{\partial u}C_{0,0}&=(T-t)\left<f(q,u)\right>_u\left(x \frac{\partial C_{0,0}}{\partial x}-C_{0,0}\right),
\end{split}
\end{equation}
in which
\begin{equation}
\left<\sigma^2(y,z)\right>_{z} \triangleq \frac{\partial}{\partial z}\left<\sigma^2(y,z)\right>, \quad \left<f(q,u)\right>_u= \frac{\partial}{\partial u} \left<f(q,u)\right>_u,
\end{equation}
and the definition of the operator $\m_1$, we can derive the
solution to (\ref{eq:c10}) as
\begin{equation}
\begin{split}
C_{0,1}(t,x,y,z)&=\frac{(T-t)^2}{2}\rho_2 \left<\sigma(y,z)\right> g(z)
\left<\sigma^2(y,z)\right>_{z} \left(x \frac{\partial}{\partial x}\left(x^2  \frac{\partial^2 C_{0,0}}{\partial x^2}\right)+\beta x^2  \frac{\partial^2 C_{0,0}}{\partial x^2}\right)
\\&-\frac{(T-t)^2}{2}g\cdot \left<\Gamma\right> \cdot \left<\sigma^2\right>_z \cdot \left(\frac{1}{2}x^2 \frac{\partial^2 C_{0,0}}{\partial x^2}+\beta x \frac{\partial}{\partial x}C_{0,0}-\beta C_{0,0}\right)
\\&-\frac{(T-t)^2}{2} \left<f\right>_u \cdot \tg \cdot \left<\tG\right>\cdot \left(x \frac{\partial C_{0,0}}{\partial x}-C_{0,0}\right)
\end{split}
\end{equation}
again using the fact that $\left<\l_2\right>$ and
$x^n\partial^n/\partial x^n$ commute for $n\in \mathbb{N}$.

 Our
approximation to the price of the call option then becomes
\begin{equation}\label{eq:app-opt-price}
\begin{split}
\widetilde{C^{\eps,\delta}} &\triangleq C_{0,0}-(T-t)\left(V_1^{\eps} x \frac{\partial}{\partial x}\left(x^2 \frac{\partial C_{0,0}}{\partial x^2}\right)+V_{2}^{\eps}x^2 \frac{\partial^2 C_{0,0}}{\partial x^2}+V_{3}^{\eps}\left(x \frac{\partial C_{0,0}}{\partial x}-C_{0,0}\right)\right)
\\&+(T-t)^2\left(V_1^{\delta} x \frac{\partial}{\partial x}\left(x^2 \frac{\partial C_{0,0}}{\partial x^2}\right)+V_{2}^{\delta}x^2 \frac{\partial^2 C_{0,0}}{\partial x^2}+V_{3}^{\delta}\left(x \frac{\partial C_{0,0}}{\partial x}-C_{0,0}\right)\right),
\end{split}
\end{equation}
in which
\begin{equation}
\begin{split}
V_{1}^{\eps}&=\sqrt{\frac{\eps}{2}}v \rho_1 \left<\sigma \phi_y\right>,
\\ V_{2}^{\eps}&=\sqrt{2 \eps}\left(\beta v \rho_1 \left<\sigma \phi_y\right>+ v \rho_1 \left<\sigma \tphi_y\right>
-\frac{ v \left<\La \phi_y\right>}{2}-\frac{ \tv \left<\tL \phi_q\right>}{2}\right),
\\ V_3^{\eps}&=-\sqrt{2 \eps}\left(v \left<\La \phi_y\right> \beta+ v \left<\La \tphi_y\right>+
\tv \left<\tL \phi_q\right> \beta+ \tv \left<\tL \tphi_q\right>\right),
\\ 2V_{1}^{\delta}&= \sqrt{\delta}\rho_2 \left<\sigma\right> g \left<\sigma^2\right>_z,
\\ 2V_{2}^{\delta}&=\sqrt{\delta}\rho_2 \cdot \left<\sigma\right> \cdot g \cdot \left<\sigma^2\right>_z \cdot \beta - \sqrt{\delta}\frac{g \cdot \left<\Ga\right>\cdot \left<\sigma^2\right>_z}{2},
\\2V_3^{\delta}&=-\sqrt{\delta}g \cdot \left<\Ga\right>\cdot \left<\sigma^2\right>_z \cdot \beta  -\sqrt{\delta}\left<f\right>_u \cdot \tilde{g} \cdot  \left<\tG\right>.
\end{split}
\end{equation}
When we calibrate (daily) to the observed option prices (\ref{eq:app-opt-price}) will be our starting point, and we will determine the 7 unknowns,
\begin{equation}\label{eq:def-lambda}
\bar{\lambda}(z,u) \triangleq \beta \left<\sigma^2(y,z)\right>+\left<f(q,u)\right>,
\end{equation}
which appears in $C_{0,0}$, $V_1^{\eps}$, $V_2^{\eps}$, $V_3^{\eps}$, $V_1^{\delta}$, $V_2^{\delta}$ and $V_3^{\delta}$ from the observed option prices.
Alternatively, we will obtain $\bar{\lambda}$ from the yield spread of defaultable bonds (either by using the yield spread of the shortest maturity bond directly or by calibrating the model implied bond price to the yield spread curve) and calibrate rest of the parameters to the observed option prices.
In either case, the average volatility term in (\ref{BS-formula}) will be calculated from the observed stock prices.

Note that when $\beta=0$ and $f=0$ we obtain the approximation formula in \cite{ronnie-timescale} as expected.
Moreover, as in
\cite{ronnie-timescale}, the accuracy of our approximation
 can be
shown to be
\begin{equation}
|C^{\eps,\delta}-\widetilde{C^{\eps,\delta}}| \leq C
(\eps|\log(\eps)|+\delta),
\end{equation}
for a constant $C>0$. The proof can be first done for smooth pay-off functions by using a higher order approximation for $C^{\eps,\delta}$ and deriving an the expression for the residual using the Feynman-Kac theorem. The proof for the call option and other non-smooth pay-offs can be performed by generalizing the regularization argument in \cite{sircar-proof}. This is where we use the boundedness assumption on the functions $f$ and $\sigma$. 

\begin{remark}
The approximation for the put option can be similarly derived. Note that the put-call parity relationship holds between the approximate call price and the approximate put price.
\end{remark}

\noindent \textbf{Note:}

 In the calibration exercises below, we will compare the performance of [the seven parameter model] in (\ref{eq:app-opt-price}) with [the model of \cite{ronnie-timescale}], which can be obtained in our framework by setting the intensity $\lambda$ in (\ref{eq:model}) to be zero. In this case stochastic volatility is the only source of the implied volatility smile/smirk.
 We will also consider the performance of two other models: [The 5 parameter model] is obtained when we set the market risk functions to be equal to zero, i.e., $\La=\tL=\Ga=\tG=0$. In this case the approximation formula becomes
 \begin{equation}\label{eq:comp-mod-1}
\begin{split}
\widetilde{C^{\eps,\delta}} &\triangleq
C_{0,0}-(T-t)\left[\tV_1^{\eps}x \frac{\partial}{\partial x} \left(x^2 \frac{\partial^2 C_{0,0}}{\partial x^2}\right)+\tV_2^{\eps}x^2 \frac{\partial^2 C_{0,0}}{\partial x^2}\right]
\\&+(T-t)^2\left[\tV_1^{\delta}x \frac{\partial}{\partial x} \left(x^2 \frac{\partial^2 C_{0,0}}{\partial x^2}\right)+ \tV_2^{\delta}x^2 \frac{\partial^2 C_{0,0}}{\partial x^2}\right],
\end{split}
\end{equation}
and the constants $\bar{\lambda}$ (the average default intensity that appears in the formula for $C_{0,0}$), $\beta$, $\tV_{1}^{\eps}$, $\tV_2^{\eps}$ and $\tV_1^{\delta}$ are to be calibrated to the observed option prices.

[The three parameter model] is obtained when we set $\La=\tL=\Ga=\tG=0$ and also take $\sigma_t=\sigma$, for some positive constant $\sigma$. In this case the approximation formula becomes
\begin{equation}\label{eq:comp-mod-2}
\widetilde{C^{\eps,\delta}} \triangleq
C_{0,0}+\left(-(T-t)V^{\eps}+(T-t)^2 V^{\delta}\right)x^2
\frac{\partial^2 C_{0,0}}{\partial x^2},
\end{equation}
and the constants $\bar{\lambda}$ (the average default intensity that appears in the formula for $C_{0,0}$), $V^{\eps}$, $V^{\delta}$, which are, again, to be calibrated to the observed option prices.
In this last model the implied volatility skew is driven only by the default intensity.

Instead of calibrating $\bar{\lambda}$ to the option prices along with the other parameters, one might prefer to obtain it from the yield spread of the corporate bond prices.

\subsection{Bond Price}

Under the zero recovery assumption, the defaultable bond price is
given by
\begin{equation}
B^{\eps,\delta}(t,Y_t,Z_t,Q_t,U_t;T)=B(t,T)\mathbb{E}\left[\exp\left(-\int_t^{T}\left(\beta \sigma^2(Y_s,Z_s)+f(Q_s,U_s)\right)dt\right)\bigg|Y_t,Z_t,Q_t,U_t\right].
\end{equation}
Here, we assume the interest rate to be constant because we are
interested in bonds with short maturities. The bond price satisfies
\begin{equation}
B^{\eps,\delta}(t,x,y,z,q,u)=:B(t,T)\tB^{\eps,\delta}(t,x,y,z,q,u),
\end{equation}
in which $\tB^{\eps,\delta}$ satisfies
\begin{equation}\label{eq:bond-pricing-pde}
\begin{split}
\tl^{\eps,\delta}\tB^{\eps,\delta}(t,x,y,z,q,u)&=0, \quad t <T,
\\ \tB^{\eps,\delta}(T,x,y,z,q,u)&=1,
\end{split}
\end{equation}
where the operator $\tl^{\eps,\delta}$ is given by
\begin{equation}
\tl^{\eps,\delta}\triangleq \frac{1}{\eps}
\l_0+\frac{1}{\sqrt{\eps}}\tl_1+\tl_2+\sqrt{\delta}\tm_1+\delta \m_2+
\sqrt{\frac{\delta}{\eps}}\m_3,
\end{equation}
in which $\l_0$, $\m_1$ and $\m_3$ are as in (\ref{eq:operators}) and
\begin{equation}
\begin{split}
\tl_1 &\triangleq -\La(y,z) v \sqrt{2} \frac{\partial}{\partial y}-\tL(q,u)\tv \sqrt{2} \frac{\partial}{\partial q},\\
\tl_2 &\triangleq \frac{\partial}{\partial t}-(\beta \sigma^2(y,z)+f(q,u))\cdot,\\
\tm_1 &\triangleq  -g(z)\Ga(y,z)\frac{\partial}{\partial z}-\tg(z)\tG(q,u)\frac{\partial}{\partial u},
\end{split}
\end{equation}

Let us obtain an expansion of the price $\tB^{\eps,\delta}$
in powers of $\sqrt{\delta}$,
\begin{equation}\label{eq:delta-expansion-bond}
\tB^{\eps,\delta}=\tB_{0}^{\eps}+\sqrt{\delta}\tB_1^{\eps}+\delta
\tB_2^{\eps}+\cdots.
\end{equation}
We will then consider the expansion of each term in
(\ref{eq:delta-expansion-bond}) in powers of $\sqrt{\eps}$, i.e.,
\begin{equation}\label{eq:expansion-eps-bond}
\tB_k^{\eps}=\tB_{0,k}+\sqrt{\eps}\tB_{1,k}+\eps \tB_{2,k}+ \cdots, \quad
\text{for all $k \in \mathbb{N}$}.
\end{equation}
The function $\tB_{0,0}(t,z,u)$ can be determined from
\begin{equation}
\left<\tl_2\right>\tB_{0,0}=0, \quad \tB_{0,0}(T,z,u)=1,
\end{equation}
as
\begin{equation}
\tB_{0,0}(t,z,u)=\exp(-\bar{\lambda}(z,u) (T-t)),
\end{equation}
in which $\bar{\lambda}$ is as in (\ref{eq:def-lambda}).
On the other hand, $\tB_{1,0}$ can be determined using
\begin{equation}
\left<\tl_2\right>\tB_{1,0}=\left<\tl_1 \l_0^{-1}\left(\tl_2-\left<\tl_2\right>\right)\right>\tB_{0,0}, \quad \tB_{1,0}(T,z,u)=0,
\end{equation}
the solution of which is
\begin{equation}
\tB_{1,0}(t,z,u)=L(z,u) (T-t) \exp \left(-\bar{\lambda}(T-t)\right),
\end{equation}
in which
\begin{equation}
L(z,u)=v \sqrt{2} \left(\beta \left<\phi_y \Lambda\right>+\left<\tphi_y \La \right>\right)+\tv \sqrt{2} \left(\beta \left<\phi_q \tL\right>+\left<\tphi_q \tL \right>\right).
\end{equation}
The function $\tB_{0,1}$ can be determined from
\begin{equation}
\left<\l_2\right>\tB_{0,1}=-\left<\tm_1\right> \tB_{0,0}, \quad \tB_{0,1}(T,x,y,z)=0.
\end{equation}
The solution of this equation can be written as
\begin{equation}
\tB_{0,1}(t,z,u)=-\tilde{L}(z,u)\frac{(T-t)^2}{2}\exp(-\bar{\lambda}(z,u)(T-t)),
\end{equation}
in which
\begin{equation}
\tilde{L}(z,u)=g \left<\Ga\right>\left(\bar{\lambda}\right)_z+\tg \left<\tG\right>\left(\bar{\lambda}\right)_q.
\end{equation}

It can be shown that
\begin{equation}
\widehat{B}^{\eps,\delta}(t,z,u;T)=B(t,T)\exp\left(\bar{\lambda}\cdot (T-t)\right)\left(1+L\cdot(T-t)-\tilde{L}\cdot\frac{(T-t)^2}{2}\right), \end{equation}
satisfies
$|\hat{B}^{\eps,\delta}-\tB^{\eps,\delta}| \leq C
(\eps+\delta)$,
for some constant $C \geq 0$, as in \cite{papa}. Since very short term maturity options are not available, we will not use this formula to calibrate $\bar{\lambda}$ from the yield spread data. Instead, in Section 3.1, we will take $\bar{\lambda}$ to be equal to the yield spread of the shortest maturity bond.

\subsection{The Approximate Implied
Volatility}\label{sec:calibration}
 The implied
volatility of the approximate pricing formula (\ref{eq:app-opt-price}), which we will denote by $I$, is implicitly defined as
\begin{equation}\label{eq:implied-vol}
 C_{\text{BS}}(x; I,r;K,T-t)=C^{\eps,\delta}(t,x,z).
\end{equation}
In order to understand how a typical implied volatility surface of our model looks like, we expand the implied volatility $I$ in terms of $\eps$ and $\delta$ as
\begin{equation}
I=I_0+\sqrt{\eps}I^{\eps}_1+\sqrt{\delta}I^{\delta}_1+\cdots
\end{equation}
We then apply the Taylor expansion formula to the left-hand-side
of (\ref{eq:implied-vol}) and use the price formula (\ref{eq:app-opt-price}) on the right-hand-side and obtain
\begin{equation}\label{eq:imp-vol-match}
\begin{split}
 &C_{\text{BS}}(x;I_0,r;K,T-t)+(\sqrt{\eps}I^{\eps}_1+\sqrt{\delta}I^{\delta}_1)
 \frac{\partial C_{\text{BS}}}{\partial \sigma
 }(x;I_0,r;K,T-t)+\cdots\\
 &=C_{0,0}-(T-t)\left(V_1^{\eps} x \frac{\partial}{\partial x}\left(x^2 \frac{\partial C_{0,0}}{\partial x^2}\right)+V_{2}^{\eps}x^2 \frac{\partial^2 C_{0,0}}{\partial x^2}+V_{3}^{\eps}\left(x \frac{\partial C_{0,0}}{\partial x}-C_{0,0}\right)\right)
\\&+(T-t)^2\left(V_1^{\delta} x \frac{\partial}{\partial x}\left(x^2 \frac{\partial C_{0,0}}{\partial x^2}\right)+V_{2}^{\delta}x^2 \frac{\partial^2 C_{0,0}}{\partial x^2}+V_{3}^{\delta}\left(x \frac{\partial C_{0,0}}{\partial x}-C_{0,0}\right)\right)+\cdots.
\end{split}
\end{equation}
Matching the order one terms, we get $I_0$ as the solution of
\begin{equation}\label{eq:defn-I-0}
C_{\text{BS}}(x;I_0,0;K B(t,T),T-t)=C_{\text{BS}}\left(x;\sqrt{\left<\sigma^2\right>},\bar{\lambda};K B(t,T),T-t\right).
\end{equation}
Note that $I_0$ is already a non-linear function of the log
moneyness, i.e. $\log(K/x)$, and of the time to maturity, i.e.
$T-t$. In fact,
\begin{equation}
x\sqrt{T-t}N'(d_1)\frac{\partial I_0}{\partial
K}=B(t,T)(N(d_2)-N(\widetilde{d}_2))<0,
\end{equation}
i.e. $I_0$ is a strictly decreasing function of $K$.
Here,
\begin{equation}
d_{1,2} \triangleq \frac{\log(x/[K B(t,T)]) \pm \frac{1}{2}I_0^2\cdot(T-t)}{I_0
\sqrt{T-t}}.
\end{equation}
The first
term in the implied volatility expansion, $I_0$, already carries the
features of a typical smile curve and captures the fact that in
the money call contracts or out of the money put contracts have
larger implied volatilities.

To match a given implied volatility curve, one will need to use
$\sqrt{\delta}$ and $\sqrt{\eps}$ terms in the expansion of the
implied volatility. Each of these terms will have one free
parameter that can be used to calibrate the model to a given
implied volatility curve. Matching the $\sqrt{\eps}$ and
$\sqrt{\delta}$ terms in (\ref{eq:imp-vol-match}), we obtain
\begin{equation}\label{eq:implied-vol-components}
\begin{split}
\sqrt{\eps}I_1^{\eps}&=-(T-t)\left(V_1^{\eps} x \frac{\partial}{\partial x}\left(x^2 \frac{\partial C_{0,0}}{\partial x^2}\right)+V_{2}^{\eps}x^2 \frac{\partial^2 C_{0,0}}{\partial x^2}+V_{3}^{\eps}\left(x \frac{\partial C_{0,0}}{\partial x}-C_{0,0}\right)\right)\left( \frac{\partial
C_{\text{BS}}}{\partial \sigma
 }\right)^{-1}  \quad \text{and} \quad
\\ \sqrt{\delta}I_1^{\delta}&= (T-t)^2\left(V_1^{\delta} x \frac{\partial}{\partial x}\left(x^2 \frac{\partial C_{0,0}}{\partial x^2}\right)+V_{2}^{\delta}x^2 \frac{\partial^2 C_{0,0}}{\partial x^2}+V_{3}^{\delta}\left(x \frac{\partial C_{0,0}}{\partial x}-C_{0,0}\right)\right)\left( \frac{\partial
C_{\text{BS}}}{\partial \sigma
 }\right)^{-1}.
\end{split}
\end{equation}
To derive these terms more explicitly we will use the derivatives of the Black-Scholes function $C_{BS}$ on the left hand side of equation (\ref{eq:defn-I-0}):
\begin{gather}
\frac{\partial C_{BS}}{\partial \sigma}=(T-t)I_0 x^2\frac{\partial^2C_{BS}}{\partial x^2}\\
\left(x\frac{\partial}{\partial x}\right)\frac{\partial}{\partial
\sigma}C_{BS}=\left(1-\frac{d_1}{I_0 \sqrt{T-t}}\right)\frac{\partial}{\partial
\sigma}C_{BS}\quad \text{and,}\\
\frac{\partial C_{BS}}{\partial \sigma}=\frac{x\exp
(-d_1^2/2)\sqrt{T-t}}{\sqrt{2\pi} }.\label{eq:vega}
\end{gather}
The corresponding derivatives of $C_{0,0}$ can be obtained by replacing $d_1$ by $\tilde{d}_1$ and $I_0$ by $\sqrt{\left<\sigma^2\right>}$. Now, we can obtain the correction terms in the implied volatilities as
\begin{equation}
\begin{split}
\sqrt{\eps}I_1^{\eps} &= -V_{1}^{\eps}\exp\left(\frac{-\tilde{d_1}^2+d_1^2}{2}\right)\left(1-\frac{\tilde{d_1}}{\sqrt{\left<\sigma^2\right>\cdot(T-t)}}\right)
-V_{2}^{\eps}\frac{1}{\left<\sigma^2\right>}\exp\left(\frac{-\tilde{d_1}^2+d_1^2}{2}\right)
\\&-V_{3}^{\eps}\sqrt{2 \pi \cdot (T-t)}\exp\left(\frac{d_1^2}{2}\right)\left(\frac{K}{x}B(t,T)N(\tilde{d}_2)\right),
\\ \sqrt{\eps}I_1^{\delta} &= (T-t)V_{1}^{\delta}\exp\left(\frac{-\tilde{d_1}^2+d_1^2}{2}\right)
\left(1-\frac{\tilde{d_1}}{\sqrt{\left<\sigma^2\right>\cdot(T-t)}}\right)
+(T-t)V_{2}^{\delta}\frac{1}{\left<\sigma^2\right>}\exp\left(\frac{-\tilde{d_1}^2+d_1^2}{2}\right)
\\&+(T-t)V_{3}^{\delta}\sqrt{2 \pi \cdot (T-t)}\exp\left(\frac{d_1^2}{2}\right)\left(\frac{K}{x}B(t,T)N(\tilde{d}_2)\right).
\end{split}
\end{equation}

\begin{figure}[h]
\begin{center}
\includegraphics[width = 0.85\textwidth,height=6cm]{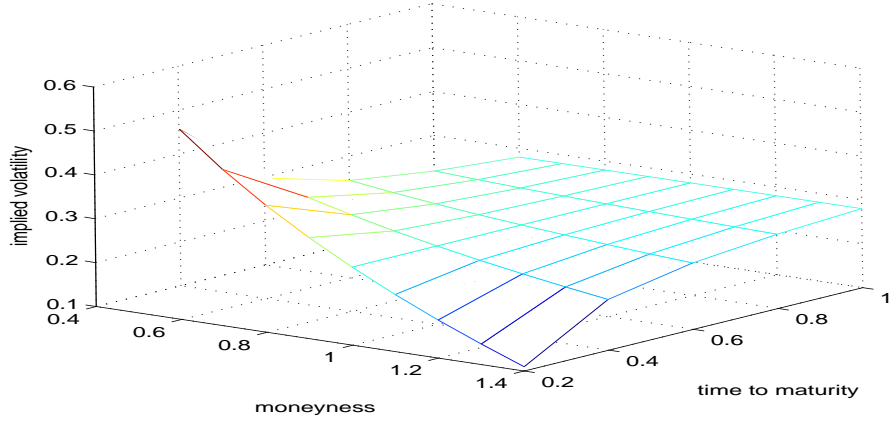}
\caption{The Implied Volatility Surface corresponding to the model of \cite{ronnie-timescale}, which is the same as (\ref{eq:comp-mod-1}) when $\bar{\lambda}=0$. Average volatility=0.2, r=0.04,  $\tV_1^{\eps}=-0.0015$ $\tV_2^{\eps}=0.001$, $\tV_1^{\delta}=-0.001$ $\tV^{\delta}=-0.001$.}
\label{fig:sfp}
\end{center}
\end{figure}

\begin{figure}[h]
\begin{center}
\includegraphics[width = 0.85\textwidth,height=6cm]{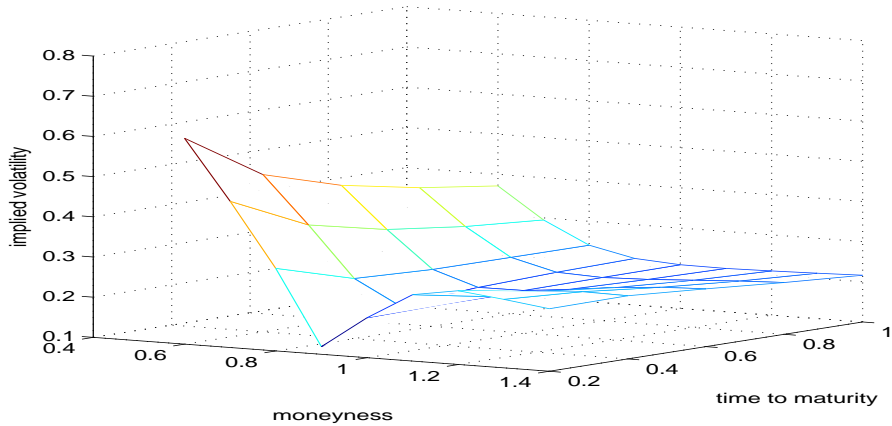}
\caption{The Implied Volatility Surface corresponding to the 7 parameter model in (\ref{eq:app-opt-price}). Average volatility=0.2, $\bar{\lambda}=0.02$, r=0.04,  $V_1^{\eps}=-0.0015$ $V_2^{\eps}=0.001$, $V_3^{\eps}=-0.005$, $V_1^{\delta}=-0.001$ $V^{\delta}=-0.001$ and $V^{\delta}_3=-0.06$.}
\label{fig:s7p}
\end{center}
\end{figure}

Typical implied volatility surfaces corresponding to the models/approximate prices in  (\ref{eq:comp-mod-2}); (\ref{eq:comp-mod-1});  and the model of \cite{ronnie-timescale}, which is the same as setting $\bar{\lambda}=0$ in (\ref{eq:comp-mod-1}); and (\ref{eq:app-opt-price}) are given respectively by Figures~\ref{fig:s3p}, \ref{fig:s5p}, \ref{fig:sfp} and \ref{fig:sfp}.  Although all of the models are able to produce the well-known properties of the implied volatility surface of the observed option prices (a-the skewedness (out of the money put options have more implied volatility then in the money ones) and b- decrease in the degree of skewedness as the time to maturity grows, c-smile (the minimum of the implied volatility is at the forward stock price $S_t B(t,T)$)) to some degree, the models in (\ref{eq:comp-mod-1}) and (\ref{eq:app-opt-price}) are able to produce surfaces with richer structures that looks more like the real implied volatility surfaces.

\section{Calibration to Data}
\subsection{Data Description}
We calibrated (daily) our model's implied volatility Ford's options over the period of 2/2/2005-1/9/2006.
The sources of data are:
\begin{itemize}
\item The stock price data is obtained from finance.yahoo.com.
\item The stock option data is from OptionMetrics under WRDS database, which is the same database \cite{carr-wu} use. It is contained in their Volatility Surface file, which contains the interpolated volatility surface using a methodology based on kernel smoothing algorithm. The file contains information on standardized options, with expirations 91, 122, 152, 182, 273, 365, 547 and 730 calender days and different strikes. There are 130 different call (put) options on each day. Exchange traded options on individual stocks are American-style and therefore the price reflects an early exercise premium. Option metrics uses binomial tree to back out the option implied volatility that explicitly accounts for this early exercise premium. We adopt the practice employed by \cite{carr-wu} for estimating the market prices of European options: We take the average of two implied volatilities at each strike and convert them into out-of-the-money European option prices using the Black-Scholes formula.

    \item The treasury yield curve data for each day is also obtained from OptionMetrics. It is derived from the LIBOR rates and settlement prices of CME Eurodollar futures. (Recall that we need the bond price $B(t,T)$ for short time to maturities $T-t$.)

    \item Ford's bond data is obtained from a Bloomberg terminal.
\end{itemize}

We will consider the performance of the three models given in (\ref{eq:app-opt-price}), (\ref{eq:comp-mod-1}) and (\ref{eq:comp-mod-2}), in fitting the implied volatility curve of the observed option data.
In the price formula (\ref{eq:app-opt-price}), we have seven parameters to calibrate: $\bar{\lambda}$, $V_i^{\eps}$, $V^{\delta}_i$, $i \in \{1,2,3\}$; on the other hand (\ref{eq:comp-mod-1}) we have five parameters to calibrate: $\bar{\lambda}$, $\tilde{V}^{\eps}_i$, $\tilde{V}^{\delta}_i$, $i \in \{1,2\};$ and finally, in (\ref{eq:comp-mod-2}) we have 3 parameters to calibrate: $\bar{\lambda}$, $V^{\eps}$, $V^{\delta}$. We will also consider the approximation formula when when we set $\bar{\lambda}=0$ (this is the approximation is the same as the one in \cite{ronnie-timescale}), and test the performance of the other models against it. The average volatility term
$\left<\sigma^2(y,z)\right>$ term that appear in our approximation formula will be estimated using the stock price data as in \cite{ronnie-timescale}.

Following the suggestion of \cite{cont} (see page 439), we perform the calibration on a particular day by minimizing
\begin{equation}\label{eq:approx-impl}
\begin{split}
\sum_i &\left(I^{\text{observed}}(T_i,K_i)-I^{\text{model}}(T_i,K_i; \text{model parameters})\right)^2
\\ &\hspace{0.5in} \approx\sum_i \frac{\left(O^{\text{observed}}(T_i,K_i)-O^{\text{model}}(T_i,K_i; \text{model parameters})\right)^2}{\text{vega}^2(T_i,K_i)},
\end{split}
\end{equation}
 in which $O^{\text{observed}}(T_i,K_i)$ is the market price of the out of the money option (put or call) with strike price $K_i$, and maturity $T_i$, where as $O^{\text{model}}(T_i,K_i; \text{model parameters})$ is the model implied price of the same option contract. The term $\text{vega}(T_i,K_i)$ is computed using (\ref{eq:vega}) with $I_0(T_i,K_i)$ as the market implied volatility.

 We will group our calibration scheme into two depending on whether the average hazard rate $\bar{\lambda}$, defined in (\ref{eq:def-lambda}) (which enters into the approximate price formulas through $C_{0,0}$ in (\ref{BS-formula}) is inferred from the yield spread of the bond prices or it is inferred as a result of calibration to the market option prices. These two groups of calibration schemes will be the topic of the following two subsections.

\subsection{When the Hazard Rate $\bar{\lambda}$ is Inferred From the Bond Yield Spread}

\begin{figure}[h]
\begin{center}
\includegraphics[width = 0.85\textwidth,height=9cm]{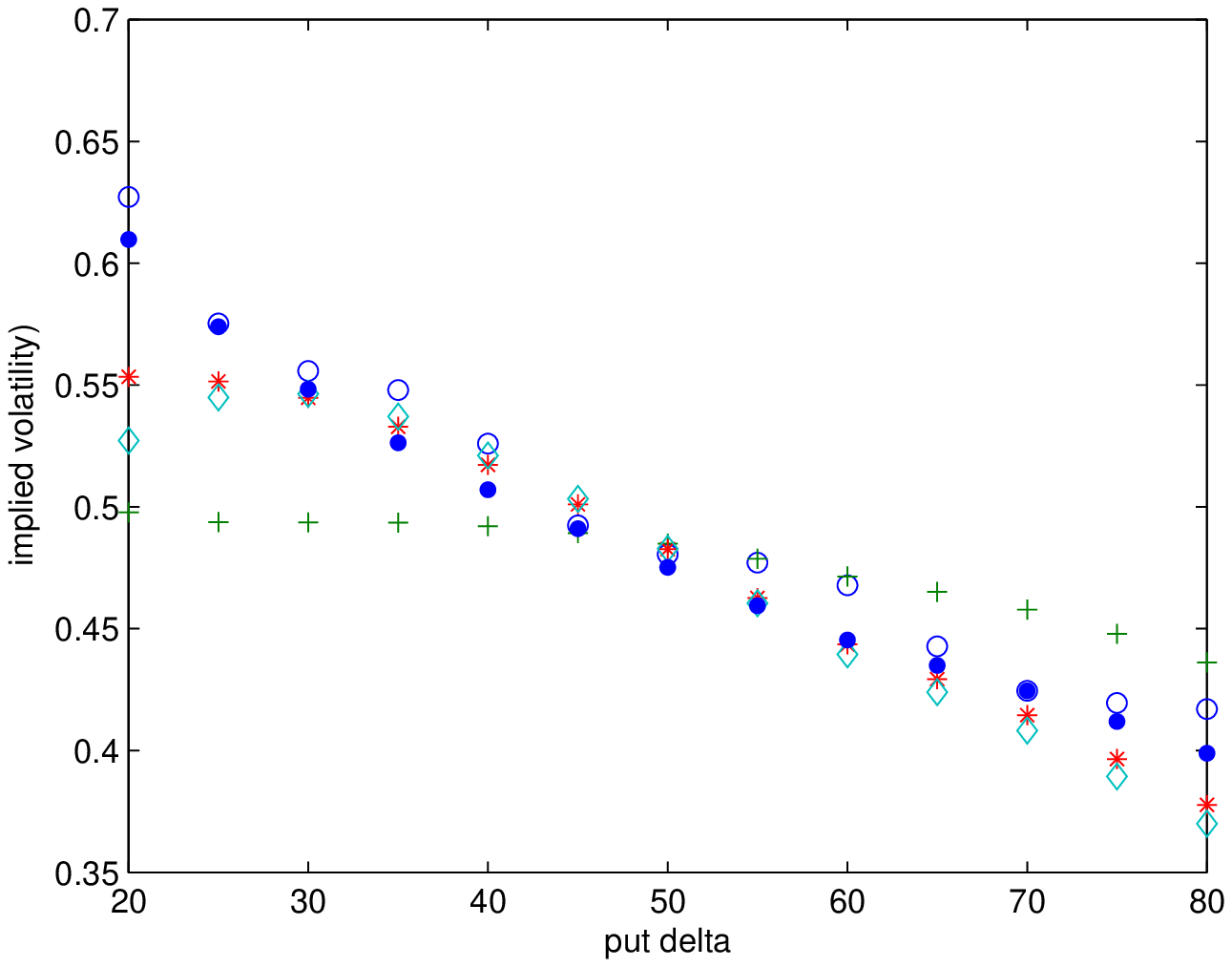}
\caption{January 9th, 2006; Maturity= 273 calender days (9 months),  Stock price=8.62, historical volatility: 29.22\%, 13 strikes, interpolated LIBOR rate=4.7771\%, $\bar{\lambda}=4.385\%$ is taken to be the yield spread of the shortest maturity bond, which matures on which matures on $1/15/2008$ (726 days). 
\newline
\textbf{Legend:}
\newline
`o': observed implied volatility;
`x': 3 parameter model implied volatility in (\ref{eq:comp-mod-2});
`*': 5 parameter model implied volatility in (\ref{eq:comp-mod-1});
full circle: 7 paramater model implied volatility in (\ref{eq:app-opt-price});
diamond: \cite{ronnie-timescale}'s implied volatility, which is obtained by setting $\bar{\lambda}=0$ in (\ref{eq:comp-mod-1}).}
\label{fig:9m}
\end{center}
\end{figure}

\begin{figure}[h]
\begin{center}
\includegraphics[width = 0.85\textwidth,height=8cm]{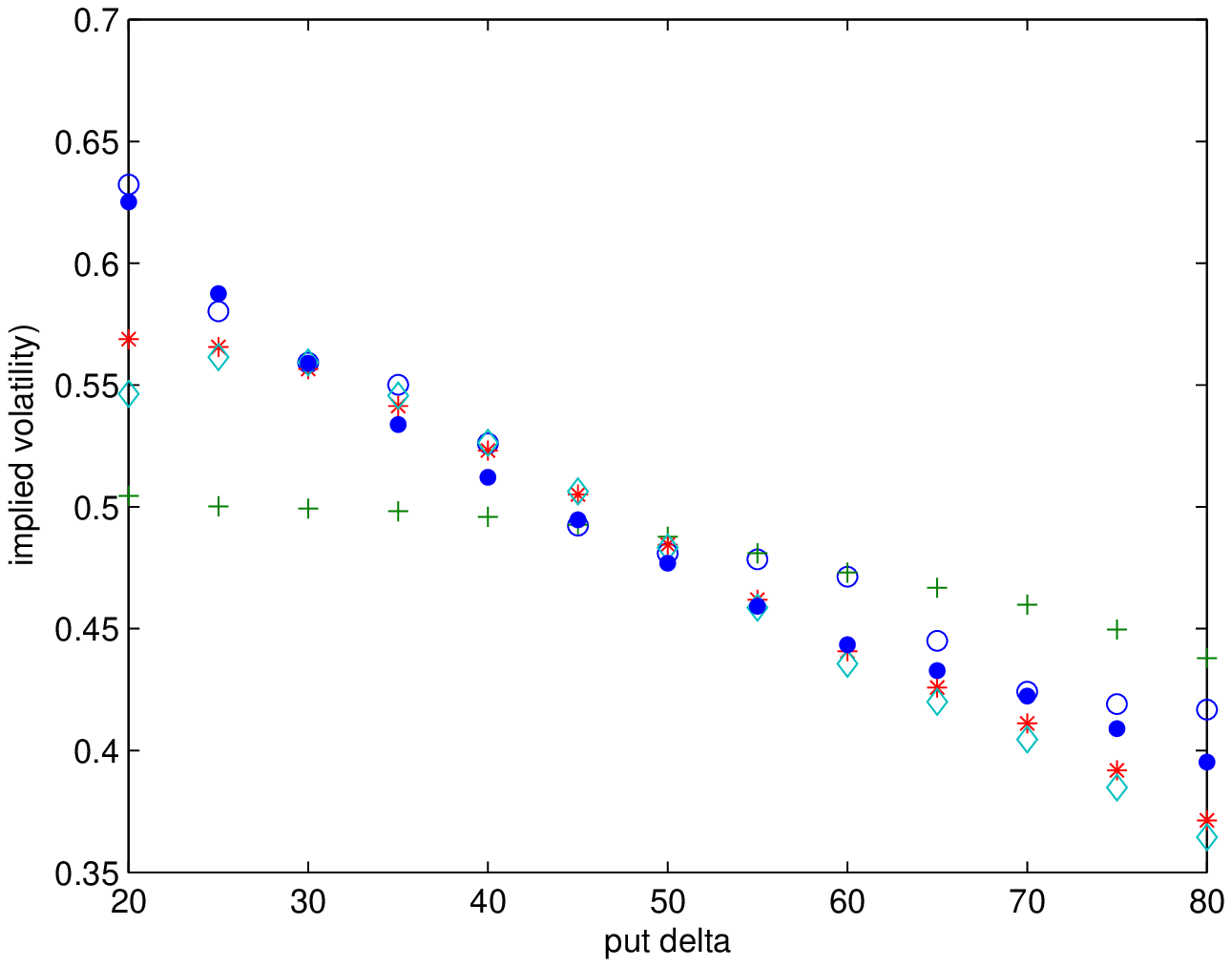}
\caption{January 9th, 2006; Maturity= 365 days,  Stock price=8.62, historical volatility: 29.22\%, interpolated LIBOR rate=4.7795\%, 13 strikes, $\bar{\lambda}=4.385\%$ is taken to be the yield spread of the shortest maturity bond, which matures on which matures on $1/15/2008$ (726 days). 
\textbf{The legend is the same as that of Fgure~\ref{fig:9m}}.}
\label{fig:1y}
\end{center}
\end{figure}

\begin{figure}[h]
\begin{center}
\includegraphics[width = 0.85\textwidth,height=9cm]{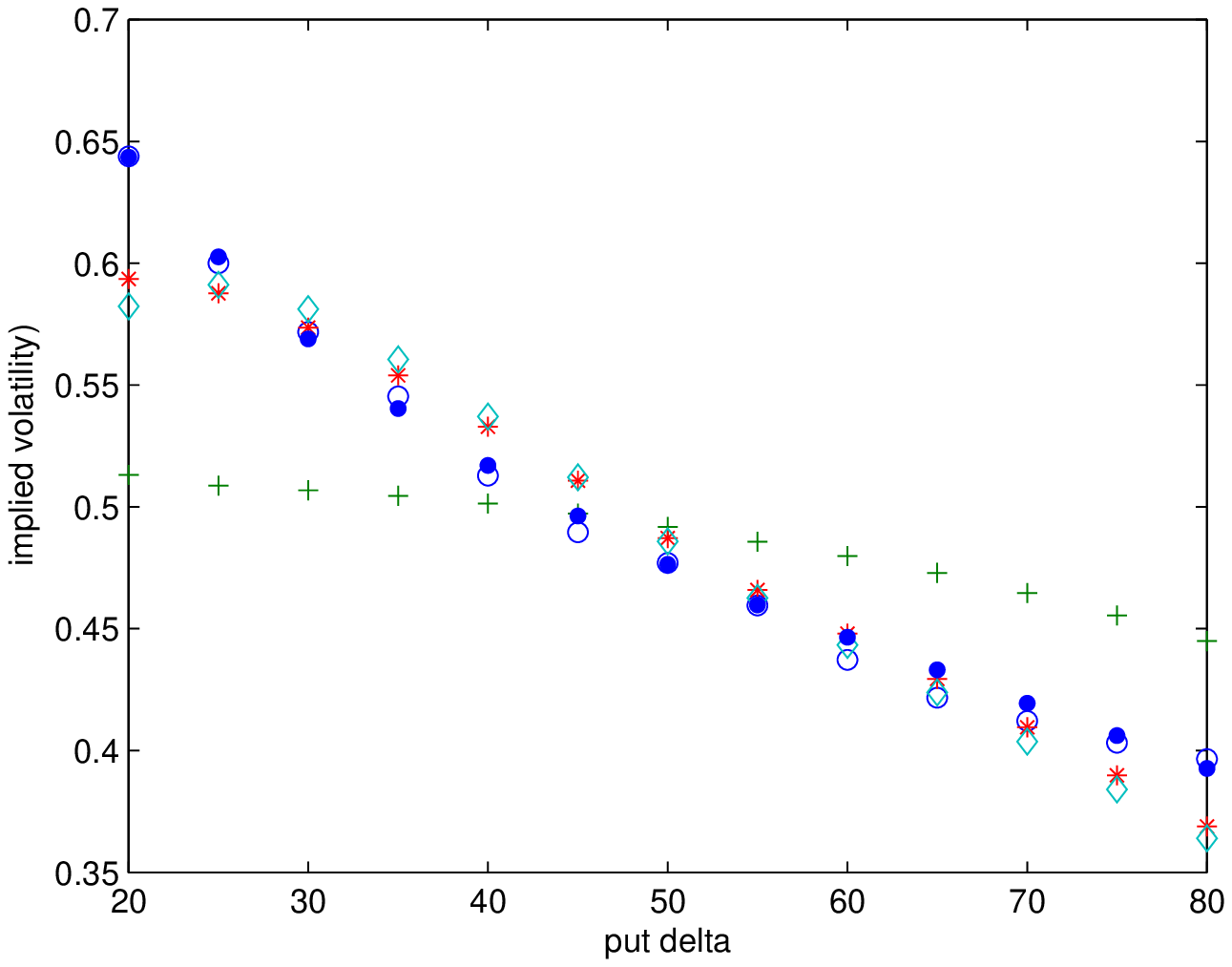}
\caption{January 9th, 2006; Maturity= 547 days (1.5 years),  Stock price=8.62, historical volatility: 29.22\%, interpolated LIBOR rate=4.7550\%, 13 strikes, $\bar{\lambda}=4.385\%$ is taken to be the yield spread of the shortest maturity bond, which matures on which matures on $1/15/2008$ (726 days). \textbf{The legend is the same as that of Fgure~\ref{fig:9m}}.}
\label{fig:1.5y}
\end{center}
\end{figure}

\begin{figure}[h]
\begin{center}
\includegraphics[width = 0.85\textwidth,height=8cm]{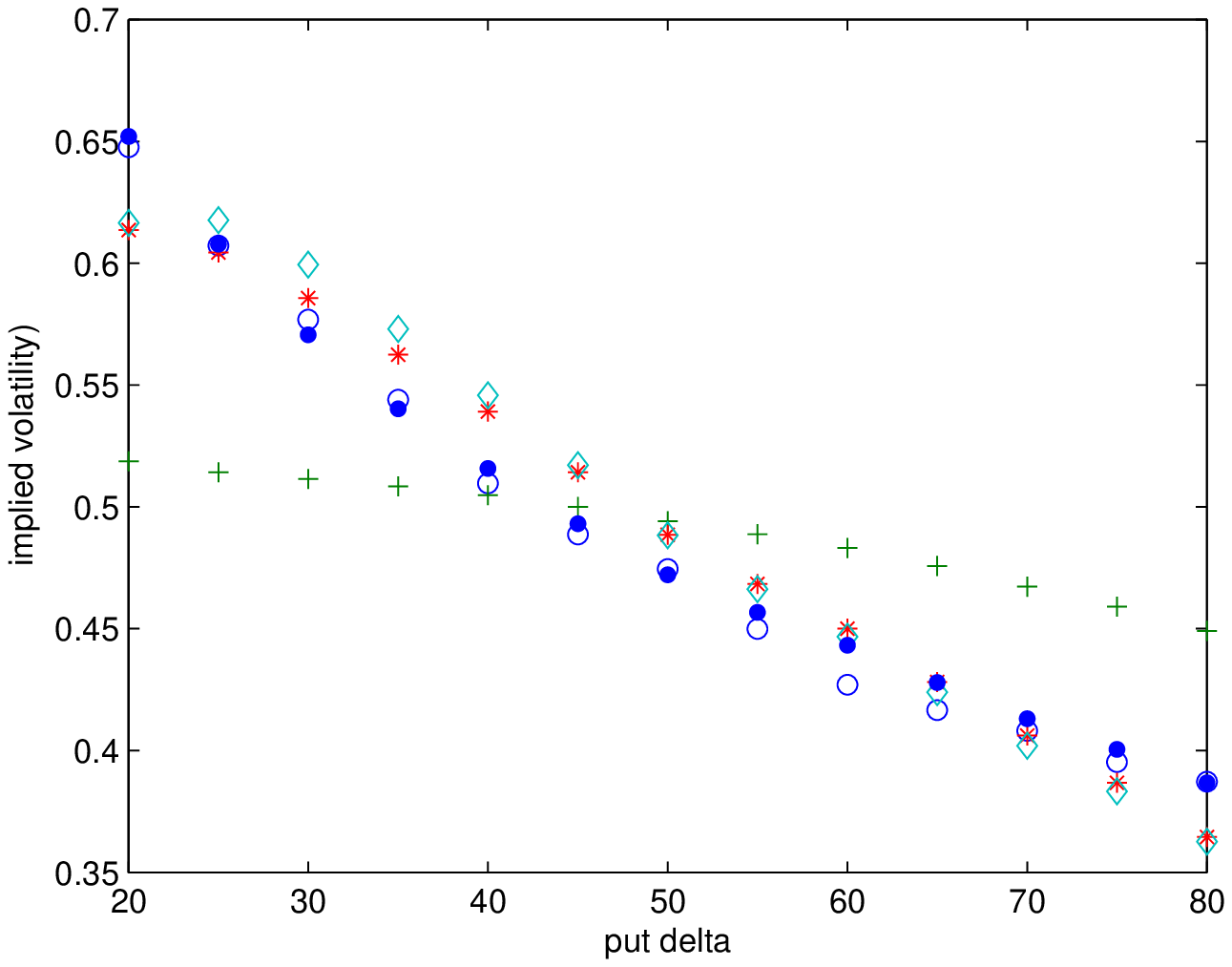}
\caption{January 9th, 2006; Maturity= 730 days (2 years),  Stock price=8.62, historical volatility: 29.22\%, interpolated LIBOR rate=4.7357\%, 13 strikes, $\bar{\lambda}=4.385\%$ is taken to be the yield spread of the shortest maturity bond, which matures on which matures on $1/15/2008$ (726 days).\textbf{The legend is the same as that of Fgure~\ref{fig:9m}}.}
\label{fig:2y}
\end{center}
\end{figure}

We fix the average hazard rate $\bar{\lambda}$ that appears in our price approximation formulas (\ref{eq:app-opt-price}), (\ref{eq:comp-mod-1}) and (\ref{eq:comp-mod-2}), to be the yield spread of the shortest maturity corporate bond (of the same company), which is 2 years in our calibration exercise. We pool the options of maturities $91, 122, 152, 182, 273, 365, 547, 730$ together and use (\ref{eq:approx-impl}) to calibrate our model. Each maturity has 13 strike prices available, therefore our data set 104 observed option prices available. After we calibrate our models and the model of \cite{ronnie-timescale}, which is obtained by setting $\bar{\lambda}=0$ in (\ref{eq:comp-mod-1}), we plot their implied volatilities against the implied volatilities of the data. These are illustrated in Figures~\ref{fig:9m}, \ref{fig:1y}, \ref{fig:1.5y} and \ref{fig:2y}. Each figure is for a different maturity.

As expected the 7 parameter model of (\ref{eq:app-opt-price}) fits the data implied volatility better. (Observe that since we fixed $\bar{\lambda}$ to be the yield spread of the shortest maturity bond there are only 6 parameters to calibrate). But the 7 parameter model does not only outperform the other models but also fits the data implied volatility almost perfectly.  7 parameter model outperforms the other models by a lot for very small and very large strikes.
Another interesting observation is that the 5 parameter model of (\ref{eq:comp-mod-1})
always outperforms (again since the $\bar{\lambda}$ is set to be the yield spread of the shortest maturity bond) the model of \cite{ronnie-timescale} which is the same as (\ref{eq:comp-mod-1}) except that the $\bar{\lambda}$ is set to zero instead of being set to the yield spread of the shortest maturity bond. This shows the significance of taking the yield spread of the corporate bonds into account in pricing options. Note that although the model of (\ref{eq:comp-mod-2}) had only two free parameters, it performed fairly well (considering that we had 104 data points available). Also note that, the performance of the 7 parameter model deteriorates as the time to maturity becomes smaller. This is because for the shorter maturity options one should use shorter term yield spreads.

\subsection{When the Hazard Rate $\bar{\lambda}$ is Inferred from Market Option Prices Directly}

In this section, we will analyze whether it would be possible to infer the risk neural expected time of default if we calibrate the models of (\ref{eq:app-opt-price}), (\ref{eq:comp-mod-1}) and (\ref{eq:comp-mod-2}) to the data implied volatility. That is along with the other parameters, if we also let $\bar{\lambda}$ be decided by the prices of the options would it be possible to say something about the yield spread of the corporate bonds.

\begin{figure}[h]
\begin{center}
\includegraphics[width = 0.85\textwidth,height=9cm]{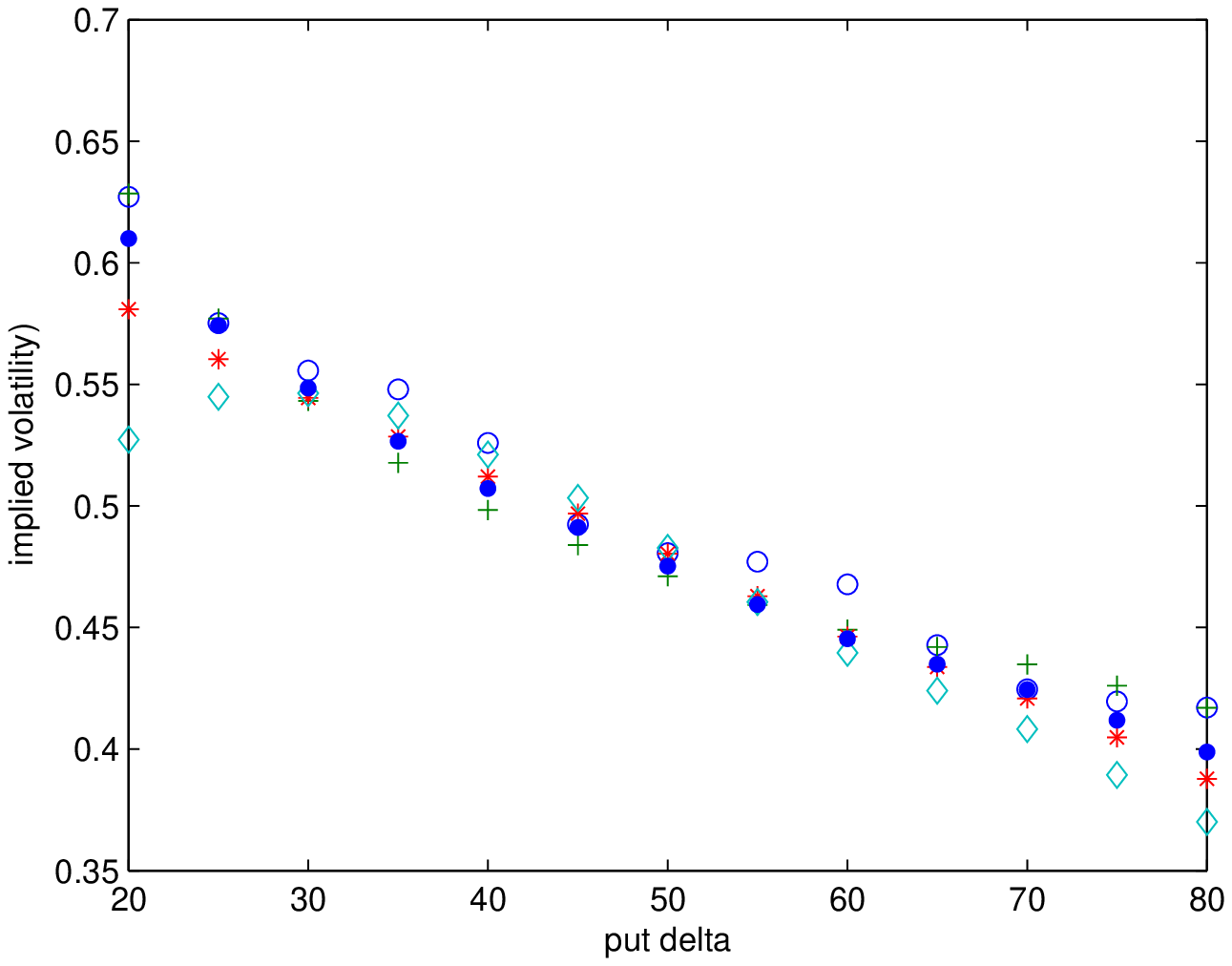}
\caption{January 9th, 2006; Maturity= 273 calender days (9 months),  Stock price=8.62, historical volatility: 29.22\%, 13 strikes, interpolated LIBOR rate=4.7771\textbf{The legend is the same as that of Fgure~\ref{fig:9m}}.}
\label{fig:9m-iter}
\end{center}
\end{figure}

\begin{figure}[h]
\begin{center}
\includegraphics[width = 0.85\textwidth,height=9cm]{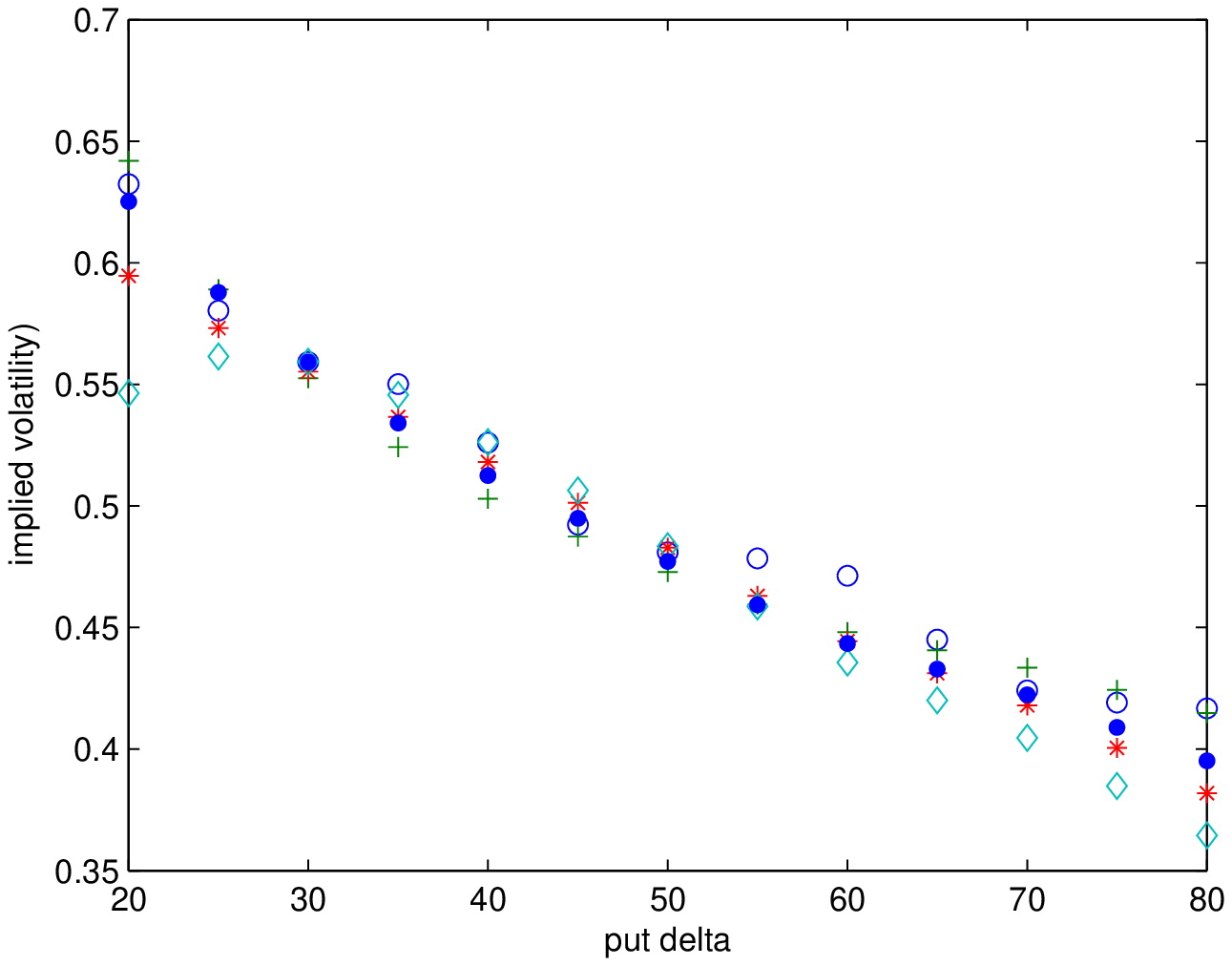}
\caption{January 9th, 2006; Maturity= 365 days,  Stock price=8.62, historical volatility: 29.22\%, 13 strikes, interpolated LIBOR rate=4.7771\%. \textbf{The legend is the same as that of Fgure~\ref{fig:9m}}.}
\label{fig:1y-iter}
\end{center}
\end{figure}

\begin{figure}[h]
\begin{center}
\includegraphics[width = 0.85\textwidth,height=9cm]{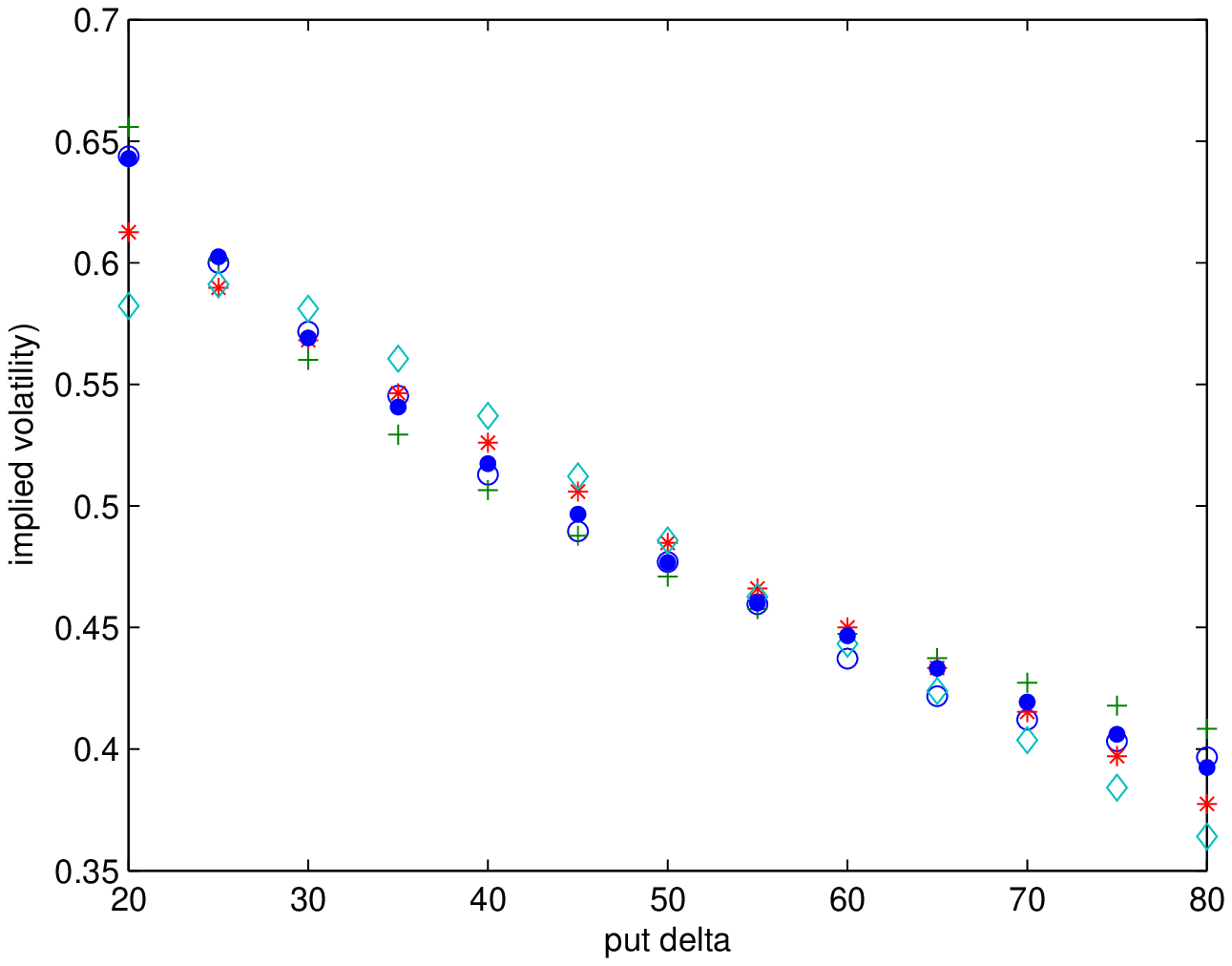}
\caption{January 9th, 2006; Maturity= 547 calender days (1.5 years),  Stock price=8.62, historical volatility: 29.22\%, 13 strikes, interpolated LIBOR rate=4.7771\%\textbf{The legend is the same as that of Fgure~\ref{fig:9m}}.}
\label{fig:1.5y-iter}
\end{center}
\end{figure}

\begin{figure}[h]
\begin{center}
\includegraphics[width = 0.85\textwidth,height=9cm]{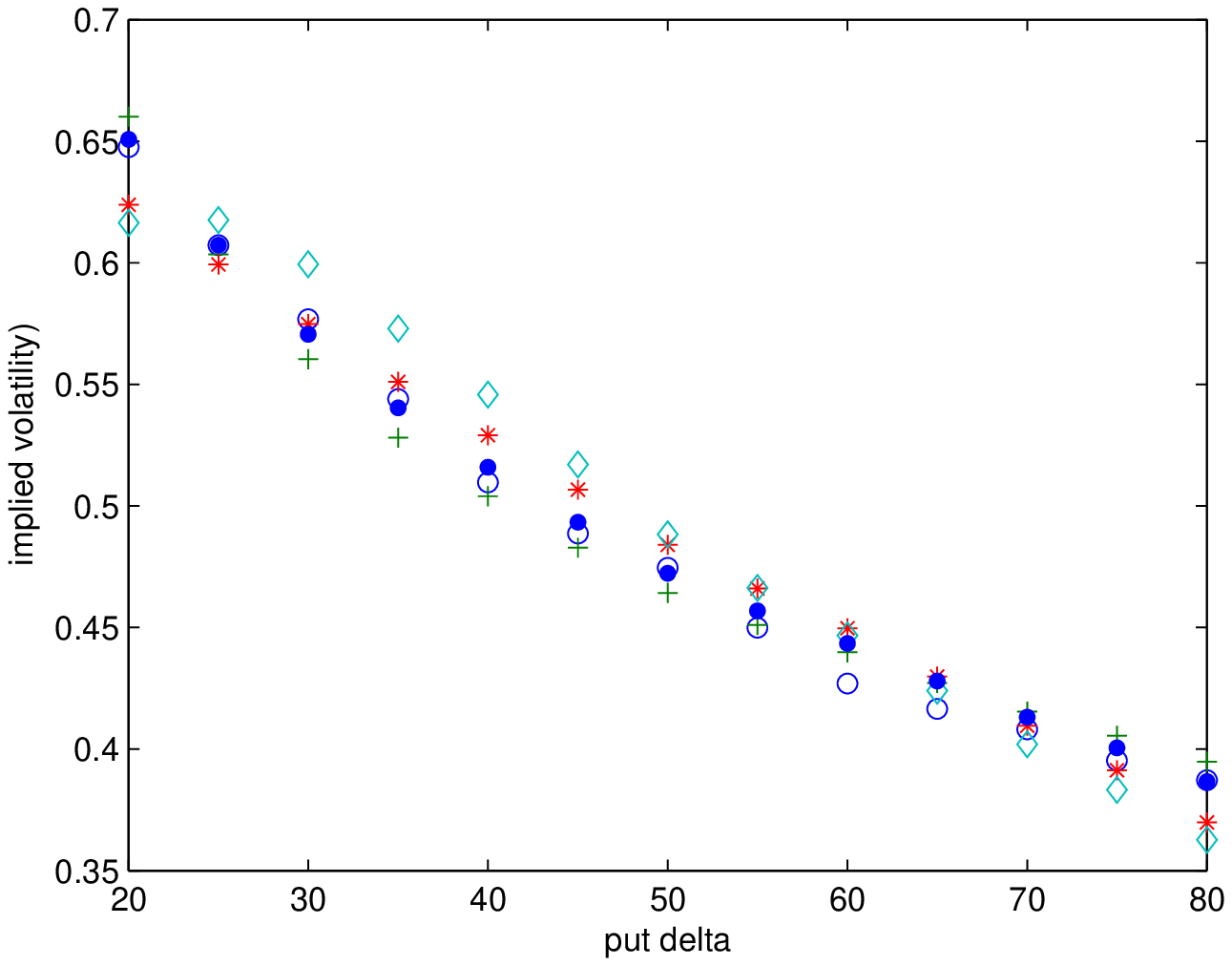}
\caption{January 9th, 2006; Maturity= 730 calender days (2 years),  Stock price=8.62, historical volatility: 29.22\%, 13 strikes, interpolated LIBOR rate=4.7771\%. \newline
`o': observed implied volatility;
`x': 3 parameter model implied volatility in (\ref{eq:comp-mod-2});
`*': 5 parameter model implied volatility in (\ref{eq:comp-mod-1});
full circle: 7 paramater model implied volatility in (\ref{eq:app-opt-price});
diamond: \cite{ronnie-timescale}'s implied volatility, which is obtained by setting $\bar{\lambda}=0$ in (\ref{eq:comp-mod-1}).}
\label{fig:2y-iter}
\end{center}
\end{figure}

\begin{figure}[h]
\begin{center}
\includegraphics[width = 0.85\textwidth,height=5cm]{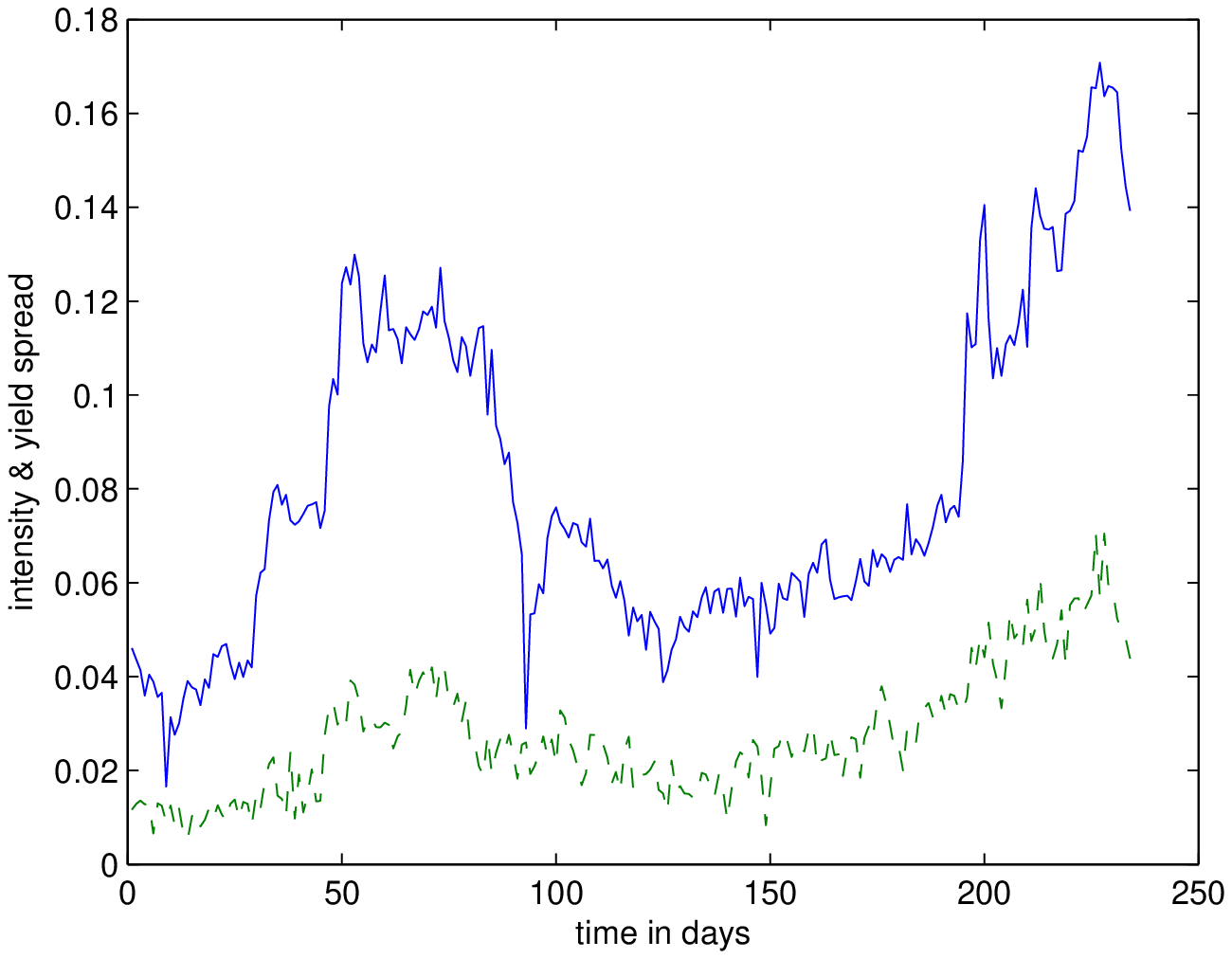}
\caption{The implied yield spread when we calibrate the 3 parameter model (\ref{eq:comp-mod-2}) to the option data
with maturities of 152, 182, 273, 365, 547, 730 (each maturity has 13 strikes) days versus
versus the yield spread of the shortest maturity corporate bond. \newline
Option implied $\bar{\lambda}$ is solid blue line, yield spread is green broken line.}
\label{fig:intensity-3p}
\end{center}
\end{figure}

\begin{figure}[h]
\begin{center}
\includegraphics[width = 0.85\textwidth,height=5cm]{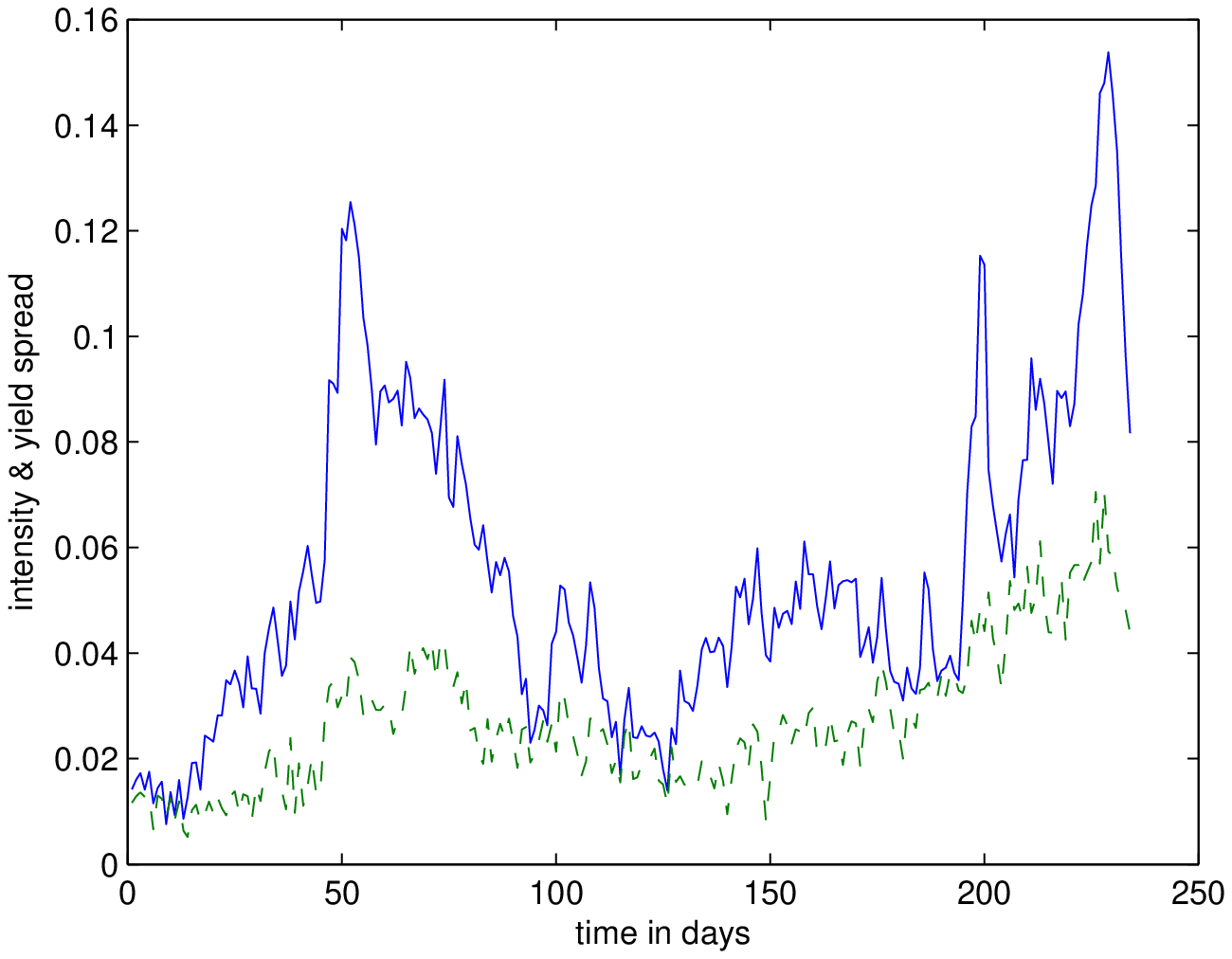}
\caption{The implied yield spread when we calibrate the 5 parameter model (\ref{eq:comp-mod-1}) to the option data
with maturities of 152, 182, 273, 365, 547, 730 (each maturity has 13 strikes) days versus
versus the yield spread of the shortest maturity corporate bond. \newline
Option implied $\bar{\lambda}$ is solid blue line, yield spread is green broken line.}
\label{fig:intensity-5p}
\end{center}
\end{figure}

\begin{figure}[h]
\begin{center}
\includegraphics[width = 0.85\textwidth,height=8cm]{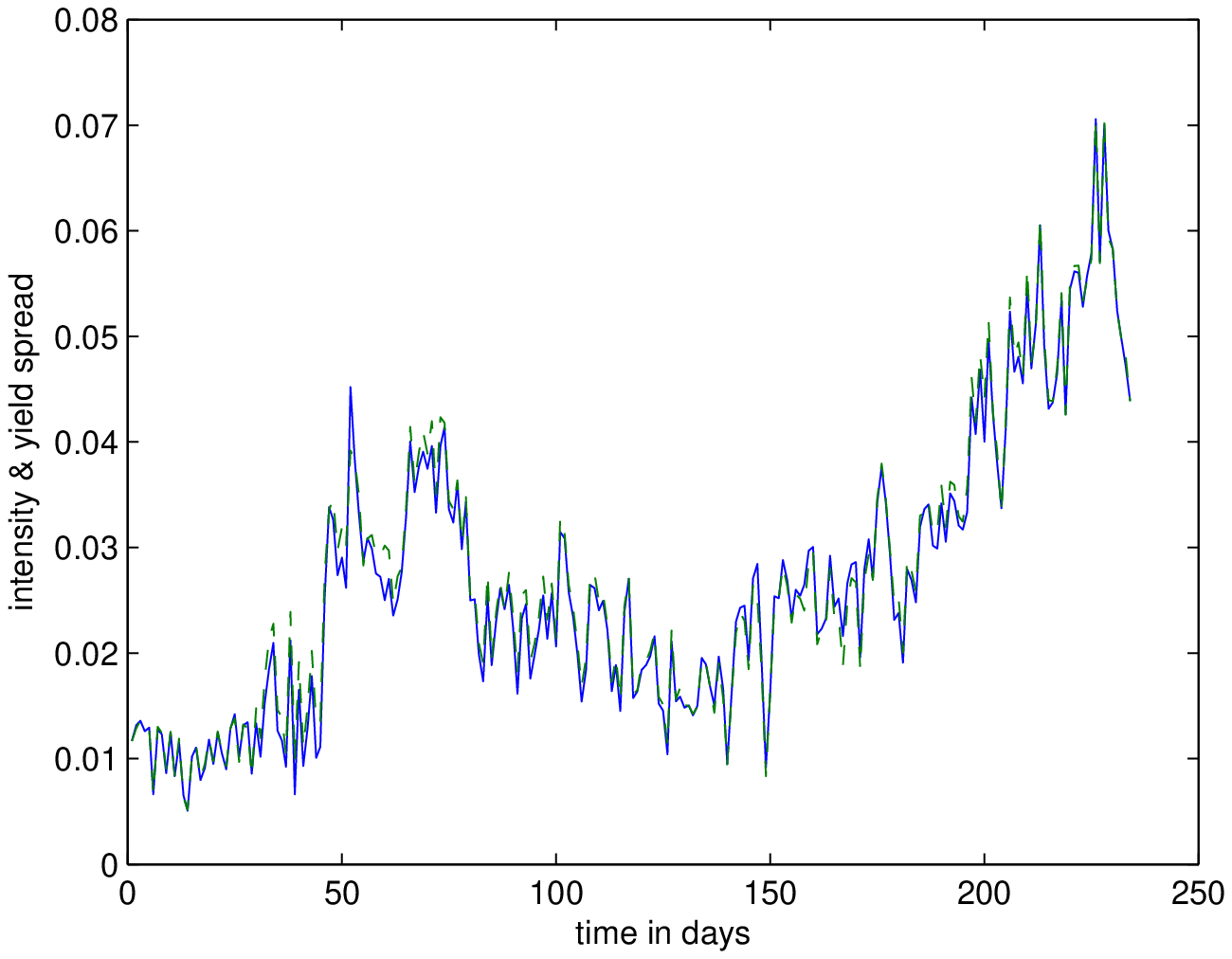}
\caption{The implied yield spread when we calibrate the 7 parameter model (\ref{eq:app-opt-price}) to the option data
with maturities of 152, 182, 273, 365, 547, 730 (each maturity has 13 strikes) days versus
versus the yield spread of the shortest maturity corporate bond. \newline
Option implied $\bar{\lambda}$ is solid blue line, yield spread is green broken line.}
\label{fig:intensity-7p}
\end{center}
\end{figure}

Since each model has one extra free variable, namely $\bar{\lambda}$, to fit to the data implied volatility, not surprisingly, the performance of each model enhances, see Figures~\ref{fig:9m-iter}, \ref{fig:1y-iter}, \ref{fig:1.5y-iter} and \ref{fig:2y-iter}. But the enhancement is not an ordinary one: all of our three models fit the data exceptionally well, even the three parameter model in (\ref{eq:comp-mod-2}). The improvement in the most advanced model is the smallest. But then when we compare the data implied yield spread versus the yield spread of the shortest maturity bond, see Figures~\ref{fig:intensity-3p}, \ref{fig:intensity-5p} and \ref{fig:intensity-7p}, the story is a little bit different. The implied yield spread, $\bar{\lambda}$ of the 3 parameter model of (\ref{eq:comp-mod-2}) is much larger than that of the yield spread of the shortest maturity corporate bond. Since, the 3 parameter model does not account for volatility, the implied yield spread is drastically severe. For the 5 parameter model the difference seems to be less severe, since the model accounts for the stochastic volatility effects, but the difference between the yield spread curves is still noticeable. On the other hand, the implied yield spread for the 7 parameter model of (\ref{eq:app-opt-price}) seems to fit the yield spread of the corporate bond almost perfectly. The difference between the 7 parameter and the 5 parameter values is that the former one accounts for the market price of intensity risk.


\newpage

\end{document}